\documentclass{aa}
\usepackage{array}
\usepackage{amsmath}
\usepackage{amssymb}
\usepackage{graphics}
\usepackage{natbib}
\bibpunct{(}{)}{;}{a}{}{,} %to follow the A&A style
\newcommand{\mean}[1]{\,\langle #1 \rangle\,}

\defcitealias{Garcia2001b}{Paper~II}

\begin{document}

%   \thesaurus{08     % A&A Section 8: Form. struct. and evolut. of stars
%              (08.09.2;  % Stars: individual
%               08.06.2;  % Stars: formation of,
%               09.10.1;  % ISM: Jets and Outflows
%               08.16.5;  % Stars: pre-main sequence
%              03.20.8);  % Techniques: spectroscopic
%              05.01.1)} % Astrometry and celestial mechanics: Astrometry
%
   \title{Atomic T~Tauri disk winds heated by ambipolar diffusion}
   \subtitle{I- Thermal structure}

   \author{P.~J.~V.   Garcia\inst{1,2,}\thanks{CAUP Support Astronomer,
              during year 2000, at the Isaac  Newton Group of Telescopes, Sta.
              Cruz de La Palma, Spain},
           J. Ferreira\inst{3},
           S. Cabrit\inst{4}
          \and
           L. Binette\inst{5}}

   \offprints{P.~J.~V. Garcia \email{pgarcia@astro.up.pt} }

   \institute{Centro de Astrof\'{\i}sica da Universidade do Porto,
              Rua das Estrelas, 4150-762 Porto, Portugal
              \and
              CRAL/Observatoire de Lyon, CNRS UMR 5574, 9 avenue
              Charles Andr\'e, 69561 St. Genis-Laval Cedex, France
              \and
              Laboratoire d'Astrophysique de l'Observatoire de Grenoble,
              BP 53, 38041 Grenoble Cedex,  France
              \and
              Observatoire de Paris,
              DEMIRM, UMR 8540 du CNRS,
              61 avenue de l'Observatoire,
              F-75014 Paris, France
              \and
              Instituto de Astronom\'\i a,
              UNAM, Ap. 70-264, 04510 D. F., M\'exico
}

   \date{Received ??/ Accepted ??}

   \authorrunning{Garcia, P., et al.}

   \titlerunning{Atomic T~Tauri disk winds heated by ambipolar diffusion}
%   \titlerunning{Thermal structure of T~Tauri disk winds}

   \abstract{
     Motivated by recent  subarcsecond resolution observations of jets
     from  T~Tauri  stars, we  extend  the  work of  \citep{Safier93a,
     Safier93b} by  computing the thermal and  ionization structure of
     self-similar,  magnetically-driven, atomic  disk winds  heated by
     ambipolar diffusion.  Improvements over his work include: (1) new
     magnetized  cold  jet solutions  consistent  with the  underlying
     accretion disk \citep{Ferreira97};  (2) a more accurate treatment
     of ionization  and ion-neutral  momentum exchange rates;  and (3)
     predictions for spatially resolved forbidden line emission (maps,
     long-slit  spectra, and  line ratios),  presented in  a companion
     paper, \cite{Garcia2001b}.\\  
     As  in  \citep{Safier93a},  we  obtain jets  with  a  temperature
     plateau around  $10^4$~K ,  but ionization fractions  are revised
     downward by a factor of  10-100. This is due to previous omission
     of thermal  speeds in ion-neutral momentum-exchange  rates and to
     different  jet  solutions.   The   physical  origin  of  the  hot
     temperature plateau  is outlined. In particular  we present three
     analytical criteria for the presence of a hot plateau, applicable
     to  any given  MHD wind  solution where  ambipolar  diffusion and
     adiabatic expansion  are the dominant heating  and cooling terms.
     We finally show that,  for solutions favored by observations, the
     jet   thermal  structure  remains   consistent  with   the  usual
     approximations  used  for   MHD  jet  calculations  (thermalized,
     perfectly    conducting,   single   hydromagnetic    cold   fluid
     calculations).
   }

   \maketitle

%
%________________________________________________________________

\section{Introduction}

Progresses in  long slit  differential astrometry techniques  and high
angular resolution  imaging from Adaptive Optics and  the Hubble Space
Telescope  have  shown  that  the  high  velocity  forbidden  emission
observed in  Classical T~Tauri Stars (CTTSs) is  related to collimated
(micro-)jets    \citep[eg.][]{Solf89,    Solf93,    Ray96,    Hirth97,
Lavalley2000,   Dougados2000a,   Bacciotti2000a}.   Although   outflow
activity is known to decrease with age \citep{Bontemps96}, CTTSs still
harbor  considerable activity  \citep[eg.][]{Mundt98} and  present the
advantage of not being embedded. It is now commonly believed that such
jets are  magnetically self-confined,  by a ``hoop  stress'' due  to a
non-vanishing poloidal  current \citep{Chan80, HN89}.  The main reason
lies  in   the  need  to  produce   highly  supersonic  unidirectional
flows. Indeed,  this requires an acceleration process  that is closely
related to the  confining mechanism. The most promising  models of jet
production  rely therefore  on the  presence of  large  scale magnetic
fields, extracting energy and mass from a rotating object. However, we
still   do   not   know   precisely   what   are   the   jet   driving
sources. Moreover, observed  jets harbor time-dependent features, with
time-scales ranging from tens  to thousands of years. Such time-scales
are  much longer  than  those  involving the  protostar  or the  inner
accretion disk. Therefore, although  the possibility remains that jets
have a non-stationary  origin \citep[eg.][]{Ouyed97c, Goodson99}, only
steady-state models will be addressed here.

Stationary    stellar     wind    models    have     been    developed
\citep[eg.][]{Sauty94},   however    observed   correlations   between
signatures of accretion and ejection  clearly show that the disk is an
essential   ingredient  in   jet   formation  \citep{Cohen89,Cabrit90,
Hartiganetal95}.   Therefore we  expect accretion  and ejection  to be
interdependent,  through the  action  of magnetic  fields.  There  are
mainly  two  classes  of   stationary  magnetized  disk  wind  models,
depending on  the radial  extent of the  wind-producing region  in the
disk. In  the first class (usually  referred to as  ``disk winds''), a
large  scale  magnetic  field  threads  the disk  on  a  large  region
\citep{BP82,   Wardle93,   Ferreira93a,   Ferreira95,   Li95,   Li96a,
Ferreira97, KB99, Casse2000a,  Casse2000b, Vlahakis2000}. Such a field
is assumed to arise from both advection of interstellar magnetic field
and local dynamo generation  \citep{Rekowski2000}. In the second class
of models (referred to as  ``X-winds''), only a tiny region around the
disk  inner edge  produces a  jet \citep{Camenzind90,  Shu94a, Shu95a,
Shu96, Lovelace99}.  The magnetic  field is assumed to originally come
from the protostar  itself, after some eruptive phase  that linked the
disk inner edge  to the protostellar magnetosphere. Note  that in both
models,  jets extract angular  momentum and  mass from  the underlying
portion  of the  disk.  However,  by construction,  ``disk-winds'' are
produced from a large spread  in radii, while ``X-winds'' arise from a
single annulus.   Apart from distinct disk physics,  the difference in
size and geometry  of the ejection regions should  also introduce some
observable jet features. Another scenario has been proposed, where the
protostar  produces  a  fast  collimated  jet  surrounded  by  a  slow
uncollimated disk wind or  disk corona \citep{Kwan88, Kwan95, Kwan97},
but such a scenario still lacks detailed calculations.

So far, all disc-driven jet calculations used a ``cold'' approximation, ie. 
negligible thermal pressure gradients. Therefore, each magnetic surface is
assumed either isothermal or adiabatic. But to  test which class of models 
is at work in T~Tauri stars, reliable observational predictions must be made
and the thermal equilibrium needs then to be solved along the flow. Such  
a difficult  task  is  still  not addressed  in  a fully 
self-consistent way,  namely by solving together the  coupled dynamics and
energy equations. Thus, no model has been able yet to predict the  gas
excitation needed to generate observational predictions.

One first possibility is to use {\em a posteriori} a simple parameterization 
for the temperature and  ionization fraction  evolution along the  flow.  This
was  done by  \cite{Shang98} and \cite{Cabrit99}  for X-winds and disk winds
respectively. These approaches are able to predict the rough jet morphology, 
but do not provide reliable flux and line profile predictions, 
since the thermal structure lacks full physical consistency. 

The  second possibility  is  to  solve the  thermal  evolution {\em  a
posteriori}, with  the difficulty  of identifying the  heating sources
(subject  to  the  constraint   of  consistency  with  the  underlying
dynamical solution). Several heating  sources are indeed possible: (1)
planar  shocks   \citep[eg.][]{Hartigan87,  Hartigan94};  (2)  oblique
magnetic shocks  in recollimating winds  \citep{Ouyed93, Ouyed94}; (3)
turbulent  mixing  layers  \cite[eg.][]{Binette99};  and  (4)  current
dissipation by  ion-neutral collisions, referred to  as {\it ambipolar
diffusion  heating} \citep{Safier93a,  Safier93b}.  A  further heating
scenario (not yet  explored in the context of MHD  jets and only valid
in   some    environments)   is   photoionization    from   OB   stars
\citep{Reipurth98,  Raga2000,  Bally2001}.    Of  all  these  previous
mechanisms only ambipolar diffusion heating allows ``minimal'' thermal
solutions,  in   the  sense  that   the  same  physical   process  ---
non-vanishing  currents  ---  is  responsible  for  jet  dynamics  and
heating.  As a consequence  no additional tunable parameter is invoked
for the  thermal description.  Furthermore,  \cite{Safier93b} was able
to   obtain  fluxes   and  profiles   in  reasonable   agreement  with
observations.  In  this paper, we extend the  work of \cite{Safier93a,
Safier93b}   by   (1)    using   magnetically-driven   jet   solutions
self-consistently computed with the underlying accretion disk, and (2)
a more  accurate treatment of  ionization using the {\tt  Mappings Ic}
code and ion-neutral momentum exchange rates which include the thermal
contribution.    In  a   companion  paper   \citep[   hereafter  Paper
II]{Garcia2001b},  we  generate  predictions  for  spatially  resolved
orbidden line emission maps, long-slit spectra, and line ratios.

This article is  structured as follows: in section~2  we introduce the
dynamical structure of the disk wind under study, and present physical
values  of the density,  velocity, magnetic  field, and  Lorentz force
along streamlines  ; in section~3  we describe the  physical processes
taken  into  account  in  the thermal  evolution  computations,  whose
results  are presented  and  discussed in  section~4. Conclusions  are
presented in section~5.   Some important derivations, dust description
and  consistency  checks of  our  calculations  are  presented in  the
appendices.

%ch2
\section{Dynamical Structure} \label{s:dyn}

\subsection{General properties}

A precise  disk wind theory must  explain how much  matter is deviated
from radial  to vertical motion, as  well as the amount  of energy and
angular momentum  carried away. This  implies a thorough  treatment of
both  the disk  interior and  its matching  with the  jets,  namely to
consider    magnetized   accretion-ejection    structures   (hereafter
MAES). The only way to solve such an entangled problem is to take into
account all dynamical terms, a task that can be properly done within a
self-similar framework.

In  this paper,  we  use the  models  of \cite{Ferreira97}  describing
steady-state, axisymmetric MAES under the following assumptions: (i) a
large  scale  magnetic  field  of  bipolar  topology  is  threading  a
geometrically thin disk; (ii) its  ionization is such that MHD applies
(neutrals are  well-coupled to the magnetic field);  (iii) some active
turbulence  inside  the  disk  produces anomalous  diffusion  allowing
matter to  cross the field  lines.  Two extra  simplifying assumptions
were  used: (iv)  jets are  assumed to  be cold,  i.e. powered  by the
magnetic  Lorentz force  only (the  centrifugal  force is  due to  the
Lorentz  azimuthal  torque), with  isothermal  magnetic surfaces  (the
midplane  temperature varying as  $T_0 \propto  r^{-1}$) and  (v) jets
carry away  all disk angular  momentum. This last assumption  has been
removed only recently by \cite{Casse2000a}.

All solutions  obtained so far  display the same  asymptotic behavior.
After  an  opening of  the  jet radius  leading  to  a very  efficient
acceleration of the plasma, the jet undergoes a refocusing towards the
axis (recollimation).  All self-similar solutions are then terminated,
most probably  producing a shock  \cite{GomezdeCastro93,Ouyed93}. This
systematic behavior could well be imposed by the self-similar geometry
itself and not be  a general result \citep{Ferreira97}.  Nevertheless,
such a shock  would occur in the asymptotic region,  far away from the
disk.   Thus,   we  can  confidently   use  these  solutions   in  the
acceleration zone,  where forbidden emission lines are  believed to be
produced \citep{Hartiganetal95}.

\subsection{Model parameters}\label{ss:dynpars}

The isothermal  self-similar MAES  considered here are  described with
three  free   dimensionless  local  parameters   \citep[see][for  more
details]{Ferreira97} and four global quantities:\\ 
(1) the disk aspect ratio
\begin{equation}
\varepsilon= \frac{h(\varpi)}{\varpi}
\end{equation}
where $h(\varpi)$ is the vertical scale height at the cylindrical 
radius $\varpi$;\\
(2) the MHD turbulence parameter
\begin{equation}
\alpha_{\rm m}= \frac{\nu_{\rm m}}{V_{\rm A}h}
\end{equation}
where $\nu_{\rm m}$ is the required turbulent magnetic diffusivity and
$V_{\rm A}$ the Alfv\'en speed  at the disk midplane; this diffusivity
allows matter  to cross field  lines and therefore to  accrete towards
the central star. It also controls the amplitude of the toroidal field
at the disk surface.\\ 
(3) the ejection index
\begin{equation}
\xi= \frac{d \ln \dot M_{\text{acc}} (\varpi)}{d \ln \varpi}
\end{equation}
which measures locally the ejection efficiency ($\xi=0$ in a standard
accretion disk), but also affects the jet opening (a higher $\xi$
translates in a lower opening);\\ 
(4) $M_*$ the mass of the central protostar;\\
(5) $\varpi_{\rm i}$ the inner edge of the MAES;\\
(6) $\varpi_{\rm e}$ the outer edge  of the MAES, a standard accretion
disk lying at greater radii.  This outer radius is formally imposed by
the amount of  open, large scale magnetic flux  threading the disk and
producing jets;\\
(7) $\dot{M}_{\rm acc}$, the disk  accretion rate fueling the MAES and
measured at $\varpi_{\rm e}$.

For our present  study, we keep only $\xi$  and $\dot{M}_{\rm acc}$ as
free parameters and  fix the values of the other  five as follows: The
disk aspect ratio was measured  by \cite{Burrows96} for HH~30 as $\sim
0.1$ so  we fix $\varepsilon =  0.1$. The MHD  turbulence parameter is
taken   $\alpha_{\text{m}}=1$   in  order   to   have  powerful   jets
\citep{Ferreira97}.   The  stellar  mass  is  fixed  at  $M_*  =0.5  \
M_{\sun}$, typical for T~Tauri stars, and the inner radius of the MAES
is set to $\varpi_{\rm i} =  $ 0.07 AU (typical disk corotation radius
for a 10 days rotation  period): inside this region the magnetic field
topology   could   be    significantly   affected   by   the   stellar
magnetosphere-disk  interaction.   The  outer radius  is kept  at
$\varpi_{\text{e}}=  1 \rm{AU}$  for  consistency with  the one  fluid
approximation (Appendix~\ref{ap:cons}) and the atomic gas description.
Regarding  atomic   consistency,  \cite{Safier93a}  solved   the  flow
evolution assuming inicially all H  bound in H$_2$.  He found H$_2$ to
completelly  dissociate  at  the  wind  base,  for  small  $\varpi_0$.
However,  after  a  critical  flow  line  footpoint  H$_2$  would  not
completelly dissociate, therefore affecting the thermal solution. This
critical  footpoint  was at  3  AU for  his  MHD  solution nearer  our
parameter space.

We  note   that  our  two   free  parameters  are  still   bounded  by
observational  constraints:  Mass  conservation relates  the  ejection
index $\xi$ to the accretion/ejection rates ratios,
\begin{equation}
2 \dot{M}_{\text{J}}\simeq \xi \dot{M}_{\text{acc}}
\ln\frac{\varpi_{\rm e}}{\varpi_{\rm i}}
\end{equation}
The observational estimates for the ratio of mass outflow rate by mass
accretion   rate   are  $\dot{M}_{\text{J}}/\dot{M}_{\text{acc}}\simeq
0.01$  \citep{Hartiganetal95}.    The  uncertainties  affecting  these
estimates   can  be  up   to  a   factor  of   10  \citep{Gullbring98,
Lavalley2000}.   The   range  of  ejection   indexes  considered  here
(0.005-0.01) is  kept compatible with Hartigan's  canonical value. The
accretion rates  $\dot{M}_{\text{acc}}$ are also kept  free but inside
the  observed  range  of  $10^{-8}\,{\rm M}_\odot  \text{yr}^{-1}$  to
$10^{-5}\,{\rm    M}_\odot    \text{yr}^{-1}$    in   T~Tauri    stars
\citep{Hartiganetal95}.

Table~\ref{tab:param} provides a list of some disk and jet parameters.
These local parameters  were constrained by steady-state requirements,
namely the smooth crossing of MHD critical points. Disk parameters are
useful to give  us a view of the physical  conditions inside the disk.
Thus, the required magnetic field $B_0$  at the disk midplane and at a
radial distance $\varpi_0$ is
\begin{equation}\label{eq:chi}
B_0  =  0.3\ \zeta  \left  ( \frac{M_*}{M_{\sun}}  \right)^\frac{1}{4}
\left  (  \frac{\dot  M_{\rm  acc}}{  10^{-7}\,M_{\sun}  \rm{yr}^{-1}}
\right)^\frac{1}{2}      \left       (      \frac{\varpi_0}{1\,\rm{AU}}
\right)^{\frac{\xi}{2} - \frac{5}{4}} \ \mbox{G} \ .
\end{equation}
The global energy conservation of a cold MAES writes
\begin{equation}
P_{\rm acc} = 2P_{\rm jet}\ +\ 2P_{\rm rad} 
\end{equation}
where $P_{\rm acc}$ is the mechanical power liberated by the accretion
flow, $P_{\rm  jet}$ the total  (kinetic, thermal and  magnetic) power
carried away by  one jet and $P_{\rm rad}$  the luminosity radiated at
one surface of  the disk. For the solutions  used, the accretion power
is given by
\begin{eqnarray}
P_{\rm acc} = \eta\  \frac{GM_* \dot M_{\rm acc}}{2\varpi_{\rm i}} &=&
0.1\eta \left ( \frac{M_*}{M_{\sun}} \right) \left ( \frac{\dot M_{\rm
acc}}  {10^{-7}\, M_{\sun}  \rm{yr}^{-1}} \right  ) \nonumber  \\  
&  &  \times \left  (  \frac{\varpi_{\rm  i}}{0.07\,{\rm AU}}  \right)^{-1}\
L_{\sun}
\end{eqnarray}
where  $L_{\sun}$ is the  solar luminosity  and the  efficiency factor
$\eta=    (\varpi_{\rm   i}/\varpi_{\rm    e})^\xi    -   (\varpi_{\rm
i}/\varpi_{\rm  e})$ depends  on  both the  local ejection  efficiency
$\xi$ and the MAES radial extent. Typical values for our solutions are
$\eta\simeq  0.9$.  The ratio  $P_{\rm jet}/P_{\rm  rad}$ is  given in
Table~\ref{tab:param}.   The jet  parameters, mass  load  $\kappa$ and
magnetic  lever  arm  $\lambda$,   have  the  same  definition  as  in
\cite{BP82}.   They are  given here  to  allow a  comparison with  the
solutions used in \citeauthor{Safier93a}'s work.

%%%%%%%%%%%%%%%%%%%%%%%%%%%%%%
\begin{table} 
\begin{center}
\begin{tabular}{ccccccc}
\hline
Solution &$\xi$ & $\zeta$ &$P_{\rm jet}/P_{\rm rad}$ & $\kappa$ &$\lambda$&
$\theta_0  (\degr)$\\ 
      \hline
A & 0.010 & 0.729 & 1.46 & 0.014 & 41.6& 50.6\\ 
B & 0.007 & 0.690 & 1.46 & 0.011 & 59.4& 52.4\\
C & 0.005 & 0.627 & 1.52 & 0.009 & 84.2& 55.4\\
\hline
\end{tabular}
\end{center}
\caption{Isothermal MAES parameters. With $\varepsilon=0.1$ and
  $\alpha_{\text{m}}=1$, the  only parameter remaining  free is $\xi$.
  Here,  the magnetic  lever  arm $\lambda$,  mass  load $\kappa$  and
  initial  jet   opening  angle  $\theta_0$  are   presented  to  ease
  comparison  with   \citeauthor{Safier93a}'s  models.  However  these
  parameters do not uniquely determine the MHD solution.}
\label{tab:param}
\end{table}
%%%%%%%%%%

%\subsection{The self-similar ansatz}
\subsection{Physical quantities along streamlines}

%fig4_____________________________________________________________________ 
\begin{figure}
  \begin{center} \resizebox{8cm}{!}{\includegraphics{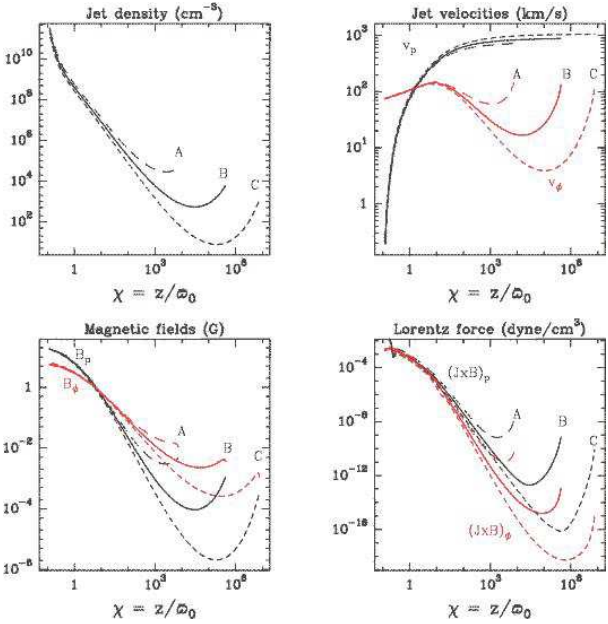}}
\end{center} 
\caption{Several wind quantities along a streamline for
model  A (long-dashed  line), B  (solid)  and C  (dashed): jet  nuclei
density $\tilde  n$, velocity, magnetic field, and  Lorentz force. For
the latter  three vectors,  poloidal components ($v_{\rm  p}$, $B_{\rm
p}$,  $(\vec J  \times  \vec B)_{\rm  p}$)  are plotted  in black  and
toroidal components ($v_\phi$ $B_\phi$, $(\vec J \times \vec B)_\phi$)
in red.  The  field line is anchored at $\varpi_0=0.1$  AU, around a 1
$M_{\sun}$  protostar,  with  an  accretion rate  $\dot  M_{\rm  acc}=
10^{-6} M_{\sun} \rm{yr}^{-1}$.}
\label{fig:adim}%
\end{figure}
%_____________________________________________________________________ 
%

In  order to obtain  a solution  for the  MAES, a  variable separation
method  has  been  used  allowing  to transform  the  set  of  coupled
partially  differential  equations  into  a set  of  coupled  ordinary
differential equations (ODEs). Hence, the solution in the $(\varpi,z)$
space is  obtained by solving  for a flow  line and then  scaling this
solution to all  space.  Once a solution is found (for  a given set of
parameters  in Section~\ref{ss:dynpars}),  the evolution  of  all wind
quantities $Q$ along any flow line is given by:
\begin{eqnarray}
Q(\varpi,z)&=&Q_0(\varpi_0) Q_\chi(\chi)
\end{eqnarray}
where  the  self-similar  variable  $\chi=  z/\varpi_0$  measures  the
position along a streamline  flowing along a magnetic surface anchored
in  the  disk at  $\varpi_0$.   In  particular,  the flow  line  shape
equation  is  given by  $\varpi(z)=  \varpi_0  \Psi(\chi)$, where  the
function  $\Psi(\chi)$  is  provided  by solving  the  full  dynamical
problem. In Fig.~\ref{fig:adim}  we plot the values of  the jet nuclei
density $\tilde  n$ and  the poloidal and  toroidal components  of jet
velocity, magnetic  field and Lorentz force.   This is done  for our 3
models,  along a  streamline with  $\varpi_0$  = 0.1~AU,  $M_{*}= 1  \
M_{\sun}$  and  $\dot{M}_{\rm   acc}  =  10^{-6}{\rm  M}_{\odot}  {\rm
yr}^{-1}$.  Values  for other  $\varpi_0$ and $\dot{M}_{\rm  acc}$ can
easily be deduced from these plots using the following scalings
\begin{eqnarray}
\tilde{n} & \propto & \dot M_{\text acc} M_\ast^{-\frac{1}{2}}
\varpi_0^{\xi - \frac{3}{2}} \nonumber \\
v & \propto & M_\ast^\frac{1}{2} \varpi_0^{-\frac{1}{2}} \nonumber \\
B & \propto & \dot M_{\text acc}^{\frac{1}{2}} M_\ast^\frac{1}{4}
\varpi_0^{\frac{\xi}{2} - \frac{5}{4}} \label{eq:scalings} \\
J & \propto & \dot M_{\text acc}^{\frac{1}{2}} M_\ast^\frac{1}{4}
\varpi_0^{\frac{\xi}{2} - \frac{9}{4}}. \nonumber
\end{eqnarray}
The  terminal poloidal velocity  is $v_{\rm{p},\infty}  \simeq \sqrt{G
M_\ast/ \xi  \varpi_0}$ so that solutions with  smaller opening angles
also   reach  smaller  terminal   velocities,  with   higher  terminal
densities.  The  point where  $v_\phi$ reaches a  minimum is  also the
point  where   the  jet  reaches   its  maximum  width  (we   call  it
recollimation point), before the jet  starts to bend towards the axis.
The numerical  solution becomes unreliable  as we move away  from this
point.   The MHD solution  is stopped  at the  super-Alfv\'enic point,
which  is reached  nearer for  higher $\xi$.   An illustration  of the
resulting ($\varpi, z$) distribution  of density $\tilde{n}$ and total
velocity  modulus for  model A  can be  found in  Fig.~1 of  Cabrit et
al. (1999).

%ch3
\section{Flow thermal and ionization processes}

\subsection{Main equations}

Under stationarity,  the thermal structure of an  atomic (perfect) gas
with density  $n$ and  temperature $T$  is given by  the first  law of
thermodynamics:
\begin{equation}\label{eq:en_cons}
P \vec{\nabla} \cdot \vec{v}\ +\ \vec{\nabla}\cdot U\vec{v} = 
\Gamma\ -\ \Lambda ,
\end{equation}
where $P=  nkT$ is the  gas pressure, $U=\frac{3}{2}nkT$  its internal
energy  density, $\vec{v}$  the total  gas velocity  and  $\Gamma$ and
$\Lambda$  are respectively  the heating  and cooling  rates  per unit
volume.  Since  during most  of the flow  the ejected gas  expands, we
call  the term $\Lambda_{\text{adia}}=  P \vec{\nabla}  \cdot \vec{v}$
the adiabatic cooling.

The gas considered here is composed of electrons, ions and neutrals of
several  atomic  species, namely  $n=  n_{\rm  e}  + \overline{n_i}  +
\overline{n_n}$ where the  overline stands for a sum  over all present
chemical elements.   We then define the density  of nuclei $\tilde{n}=
\overline{n_i} + \overline{n_n}$ and  the electron density $n_{\rm e}=
f_{\rm e}  \tilde{n}$.  Correspondingly, the  total velocity $\vec{v}$
appearing  in   Eq.~(\ref{eq:en_cons})  must  be   understood  as  the
barycentric velocity.  As usual in one-fluid approximation, we suppose
-- and  verify it  in  section~\ref{ap:cons:dyn} --  all species  well
coupled (through collisions), so  that they share the same temperature
$T$.  We also  assume that no molecule formation  occurs, so that mass
conservation requires
\begin{equation} \label{eq:mass}
\vec{\nabla}\cdot \tilde{n} \vec{v} = 0 \ .
\end{equation}
Under stationarity, the gas species ionization state evolves according
to the rate equations,
\begin{eqnarray}\label{eq:rate}
\frac{Df_{A^i}}{Dt} &=& %\frac{\partial f_{A^i}}{\partial t} + 
\vec{v}\cdot \vec{\nabla} f_{A^i}  = \frac{R_{A^i}}{\tilde n} ,\\
\frac{Df_{\rm e}}{Dt} &=& %\frac{\partial f_{\rm e}}{\partial t} + 
\vec{v}\cdot \vec{\nabla} f_{\rm e}
= \sum_{A} \sum_{i=1}^{i \leqslant N_A} i \, \frac{R_{A^i}}{\tilde n} ,
\end{eqnarray}
subjected to the elemental conservation constraint,
\begin{eqnarray} \label{eq:el_cons}
\sum_{i=0}^{i \leqslant N_A} R_{A^i} = 0 .
\end{eqnarray}
In the  above equations $f_{A^i}= n_{A^i}/\tilde n$  is the population
fraction of the chemical element $A$, $i$ times ionized, $R_{A^i}$ the
rate of  change of  that element and  $N_A$ is its  maximum ionization
state.  Note that $R_{A^i}$ is a function of all the species densities
$n_{A^i}$,  the temperature  and  the radiation  field.   In order  to
obtain the gas temperature and  ionization state, we must solve energy
equation coupled  to the ionization evolution.  This  task requires to
specify  the  ionization  mechanisms   as  well  as  the  gas  heating
($\Gamma$)   and    cooling   ($\Lambda$)   processes    (other   than
$\Lambda_{\text{adia}}$).

\subsection{Ionization Evolution}

\subsubsection{The {\tt Mappings  Ic} code}

We   solve   the   gas   ionization   state   (Eqs.~\ref{eq:rate}   to
\ref{eq:el_cons})    using   the   {\tt    Mappings   Ic}    code   --
\citet{Binette85, Binette87,  Ferruit97}.  This code  considers atomic
gas composed by the chemical elements H,  He, C, N, O, Ne, Fe, Mg, Si,
S, Ca, Ar. We also added  Na (whose ionization evolution is not solved
by  {\tt  Mappings Ic}),  assuming  it  to  be completely  ionized  in
\ion{Na}{ii}.  Hydrogen and Helium are treated as five level atoms.

The   rate   equations   solved   by   {\tt   Mappings   Ic}   include
photoionization, collisional  ionization, secondary ionization  due to
energetic   photoelectrons,   charge   exchange,   recombination   and
dielectronic    recombination.     This    is   in    contrast    with
\citeauthor{Safier93a},  who assumed a  fixed ionization  fraction for
the heavy elements  and solved only for the  ionization evolution of H
and He, considering two levels for H and only the ground level for He.

The  adopted abundances are  presented in  Table~2.  In  contrast with
\citeauthor{Safier93a}, we  take into account  heavy element depletion
onto      dust     grains      (see      section~\ref{ss:dust}     and
Appendix~\ref{ap:dust}) in the dusty region of the wind.

%%%%%%%%%%%%%%%%%%%%%%%%%%%%%%
\begin{table}
\begin{center}
\begin{tabular}{ >{\scriptsize}c  >{\scriptsize}c  >{\scriptsize}c
                 >{\scriptsize}c } 
\hline
Element &Z$_{\odot}$ &Z$_{\text{DL84}}$ &Z$_{\text{d}}$ \\ \hline
H  &1.0 & &\\
He &1.0 (-1) & & \\
C  &3.55(-4)&3.0 (-4) &2.17(-4) \\
N  &9.33(-5)& & 1.39(-5)\\
O  &7.41(-4)&1.36(-4) &3.39(-4)  \\
Ne &1.23(-4)& &  \\
Fe &3.24(-5)&3.01(-5) &3.22(-5) \\
Mg &3.80(-5)&3.71(-5) &3.69(-5)  \\
Si &3.55(-5)&3.33(-5) & 3.37(-5) \\
S  &1.86(-5)& &  \\
Ca &2.19(-6)&  &2.19(-6) \\
Ar &3.63(-6)& &2.43(-6) \\
Na &2.04(-6)& &1.81(-6)\\
\hline
\end{tabular}
\end{center}
\caption{
Abundances  by number of  various elements  with respect  to Hydrogen.
The  notation 3.55(-4)  means  $3.55\times10^{-4}$.  Column  Z$_\odot$
gives     solar    abundances,    from     \cite{Savage96}.     Column
Z$_{\text{DL84}}$ elemental abundances locked in grains for a MRN-type
dust  distribution,  from  \cite{DraineLee84}.  Column  Z$_{\text{d}}$
gives the  abundances locked  in grains in  the diffuse  clouds toward
$\zeta$ Ophiuchi \citep{Savage96}), computed using the solar gas phase
abundances  from the  same authors.   The depleted  abundances adopted
here are $Z=Z_\odot-Z_{\text{d}}$.}
\end{table}
%%%%%%%%%%

\subsubsection{Photoionizing radiation Field}\label{ss:radiation}

For  simplicity, the central  source radiation  field is  described in
exactly the  same way  as in \citeauthor{Safier93a}  and we  refer the
reader to the expressions  (C1-C10) presented in his Appendix~C.  This
radiation  field is  diluted with  distance  but is  also absorbed  by
intervening wind material ejected at smaller radii.

We treat the radiative transfer as a simple absorption of the diluted
central source, namely
\begin{eqnarray}
%4 \pi J_\nu = \pi F_\nu(r) \exp(-\tau_\nu(r)) \label{eq:rad}
4 \pi J_\nu(r, \theta) = \frac{ L_\nu (\theta)}{4 \pi r^2} e^{- \tau_\nu(r,
  \theta)}\label{eq:rad},
\end{eqnarray}
where $J_\nu(r,\theta)$ is the local mean monochromatic intensity at a
spherical  radius  $r$ and  angle  with  the  disk vertical  $\theta$,
$L_\nu(\theta)$ is  the emitted luminosity  of both star  and boundary
layer and  $\tau_\nu(r,\theta)$ the optical depth  towards the central
object.

We now address the question of  optical depth.  In our model, the flow
is  hollow, starting  from a  ring located  at the  inner  disk radius
$\varpi_{\rm i}$  and extending to the outer  radius $\varpi_{\rm e}$.
The jet  inner boundary is  therefore exposed to the  central ionizing
radiation,  which  produces  then  a  small layer  where  hydrogen  is
completely photoionized.   The width $\Delta  r$ of this layer  can be
computed by equating the number of emitted H ionizing photons, $Q({\rm
H}_0) = \frac{1}{2} \int_{\nu_0}^{\infty} L_\nu  d\nu / h \nu$, to the
number of recombinations in this layer, $n_{\rm H}^2\alpha_{\rm B}(T)2
\pi r^2\Delta  r$ for our geometry.   We found that $\Delta  r \ll r$,
and  thus assume  that all  photons capable  of ionizing  Hydrogen are
exhausted within  this thin  shell.  Furthermore, there  is presumably
matter in the inner  ``hollow'' region, so the previous considerations
are upper limits.

For the heavy elements, photoionization optical depths are negligible,
due to the much smaller abundances, and are thus ignored.  The opacity
$\tau_\nu$  is  assumed  to  be  dominated  by  dust  absorption  (see
Appendix~\ref{ap:dust}   for  details).    Dust  will   influence  the
ionization  structure at  the base  of the  flow, where  ionization is
dominated by heavy elements.

To summarize, the adopted radiation field is a central source absorbed
by dust, with a cutoff at and above the Hydrogen ionization frequency.

\subsubsection{Dust properties and gas depletion}\label{ss:dust}

\citeauthor{Safier93a}  showed that  if dust  exists inside  the disk,
then the wind  drag will lift the dust thereby  creating a dusty wind.
Our   wind   shares   the   same   property.   We   model   the   dust
(Appendix~\ref{ap:dust})  as  a   mix  of  graphite  and  astronomical
silicate, with  a MRN size distribution  and use for  the dust optical
properties  the tabulated  values of  \citet{DraineLee84, DraineMal93,
LaorDraine93}.  For  simplicity we assumed the dust  to be stationary,
in  thermodynamic equilibrium  with  the central  radiation field  and
averaged all dust quantities by the MRN size distribution.

In addition, we take into account depletion of heavy elements into the
dust phase.  This effect was not considered by \citeauthor{Safier93a}.
In Table~2 we present the dust phase abundances needed to maintain the
MRN distribution (\citeauthor{DraineLee84}),  and our adopted depleted
abundances, taken  from observations of diffuse  clouds toward $\zeta$
Ori \citep{Savage96}.   These are more  realistic, although presenting
less depletion of  carbon than required by MRN.   Depletion has only a
small  effect on  the calculated  wind thermal  structure, but  can be
significant when  comparing to observed line ratios  based on depleted
elements.

\subsection{Heating \& Cooling Mechanisms}

\subsubsection{MHD heating}

The dissipation of electric currents $\vec J$ provides a local heating
term  per  unit volume  $\Gamma_{\rm  MHD}= \vec  J  \cdot  (\vec E  +
\frac{\vec v} {c} \times \vec B)$, where $\vec E$ and $\vec B$ are the
electric and  magnetic fields ,  $\vec{v}$ the fluid velocity  and $c$
the  speed of  light. In  a  multi-component gas,  with electrons  and
several ion and neutral species, the generalized Ohm's law writes
\begin{eqnarray}
\vec E + \frac{\vec v}{c} \times\vec B &=& \overline{\eta} \vec{J} 
- \left(\frac{\overline{\rho_n}}{\rho}\right)^2 \frac{\frac{1}{c^2}
(\vec J \times \vec B)\times \vec B}{\overline{m_{in} n_i \nu_{in}}}
 \nonumber \\
& & - \frac{\vec{\nabla} P_{\rm e}}{e n_{\rm e}} + \frac{\vec J \times\vec B}{e n_{\rm e}c},
\end{eqnarray}
where $\eta$ is the  fluid electrical resistivity, $\rho$ and $\rho_n$
are  the  total  and   neutral  mass  density,  $m_{in}$  the  reduced
ion-neutral  mass,   $n_i$  the  ionic  density   and  $\nu_{in}$  the
ion-neutral collision  frequency.  The overline stands for  a sum over
all  chemical  elements relevant  to  a  given  quantity.  These  last
quantities depend on the gas ionization state, the temperature and the
momentum  exchange  rate  coefficients.   The reader  is  referred  to
Appendix~\ref{ap:dissipa}  for the  expressions of  these coefficients
and the uncertainties affecting them.

The first  term appearing  in the right  hand side of  the generalized
Ohm's  law is the  usual Ohm's  term, while  the second  describes the
ambipolar  diffusion, the  third  is  the electric  field  due to  the
electron  pressure and the  last is  the Hall  term. This  last effect
provides no net dissipation in  contrary to the other three.  It turns
out  that the  dissipation due  to  the electronic  pressure is  quite
negligible  and has  been therefore  omitted (Appendix~\ref{ap:cons}).
Thus, the MHD dissipation writes
\begin{eqnarray}\label{eq:dissipa}
\Gamma_{\text{MHD}} &=& \overline{\eta}J^{2} +
\Big(\frac{ \overline{\rho_n} } {\rho }\Big)^2  
\frac{\frac{1}{c^2}\bigl\|  \vec{J}\times \vec{B}
\bigr\|^{2}}{\overline{m_{in} n_i   \nu_{in}}}\notag\\
&\equiv&   \Gamma_{\text{Ohm}}   + \Gamma_{\text{drag}}.
\end{eqnarray}
The  first term,  Ohmic heating  $\Gamma_{\text{Ohm}}$,  arises mainly
from  ion-electron drag. The  second term  is the  ambipolar diffusion
heating $\Gamma_{\text{drag}}$ and is  mainly due to ion-neutral drag.
This   last    term   is    the   dominant   heating    mechanism   in
\citeauthor{Safier93a}'s disk wind models, as well as in ours.

An important  difference with  \citeauthor{Safier93a} is that  we take
into  account thermal  speeds  in ion-neutral  momentum exchange  rate
coefficients. This increases  $\nu_{in}$, and results in significantly
smaller ionization fractions (Sect.~4.8).

\subsubsection{Ionization/recombination cooling}

Both   collisional  ionization   cooling   $\Lambda_{\text{col}}$  and
radiative  recombination  cooling  $\Lambda_{\text{rec}}$ effects  are
taken into account by {\tt Mappings Ic}. These terms are given by,
\begin{eqnarray}
\Lambda_{\text{col}}  &=&   \sum_{A}  \sum_{i=0}^{i<N_A}  R^c_{A^i}  I_{A^i}\\
\Lambda_{\text{rec}}  &=& \sum_{A}  \sum_{i=0}^{i<N_A}  
n_{\rm e} n_{A^{i+1}} k T\beta_{\rm B}(A^i)
\end{eqnarray}
where $R^c_{A^i}$  is the  collisional ionization rate,  $I_{A^i}$ the
ionization  energy and $\beta_{\rm  B}(A^i)$ is  the case  B radiative
recombination rate to the ionization state $A^i$.

These ionization/recombination effects, taken  into account in part by
\citeauthor{Safier93a}, are in general smaller than adiabatic and line
cooling.

\subsubsection{Photoionization heating and radiative cooling}

Photoionization  by the  radiation field,  not taken  into  account by
\citeauthor{Safier93a},   provides   an   extra  source   of   heating
$\Gamma_{\text{P}}$.  This  term,  which  is  also  computed  by  {\tt
Mappings Ic}, is given by
\begin{eqnarray}
\Gamma_{\text{P}} = \sum_A \sum_{i=0}^{i<N_A} n_{A^i}
\int_{\nu_0^{A^i}}^{\infty} \frac{4 \pi J_\nu}{h \nu}
a_\nu^{A^i} (h\nu - h \nu_0^{A^i}) d\nu
\end{eqnarray}
where $\nu_0^{A^i}$  is the threshold frequency for  ionization of the
chemical element $A$, $i$ times ionized, $4\pi J_\nu$ is the radiation
flux of  the central source  (described in section~\ref{ss:radiation})
and $a_\nu^{A^i}$  the photoionization cross-section.  We  found it to
be the dominant  heating source at the base of the  flow, at the inner
radii and for high accretion rates.

Collisionally excited lines  provide a very efficient way  to cool the
gas,  thanks to an  extensive set  of resonance  and inter-combination
lines,   as  well   as  forbidden   lines.   This   radiative  cooling
$\Lambda_{\rm rad}$  is computed by  {\tt Mappings Ic} by  solving for
each  atom the  local statistical  equilibrium, and  will allow  us to
compute  emission   maps  and   line  profiles  for   comparison  with
observations  \citepalias[see][]{Garcia2001b}.   We include  cooling by
hydrogen lines, $\Lambda_{\rm rad}({\rm H})$, in particular H$\alpha$,
which could not be  computed by \citeauthor{Safier93a} (two-level atom
description).

\subsubsection{Other minor heating/cooling mechanisms}

Several processes, also computed by  {\tt Mappings Ic}, appeared to be
very small and  not affecting the jet thermal  structure. We just cite
them here for completeness:  free-free cooling and heating, two-photon
continuum and Compton scattering.

We  ignored thermal  conduction, which  could be  relevant  along flow
lines, the  magnetic field reducing the gas  thermal conductibility in
any other direction.  Also ignored  was gas cooling by dust grains and
heating by cosmic rays. We checked {\em a posteriori} that these three
terms     indeed    have     a     negligible    contribution     (see
Appendix~\ref{ss:ign_h}).

\subsection{Numerical resolution}

In our study,  flow thermodynamics are decoupled from  the dynamics --
cold jet  approximation. The previous  equations (\ref{eq:en_cons}) to
(\ref{eq:el_cons}) can  then be integrated  for a given  flow pattern.
The dynamical  quantities (density, velocity and  magnetic fields) are
given by the cold  MHD solutions presented in Section~\ref{s:dyn}. For
the  steady-state, axisymmetric, self-similar  MHD winds  under study,
any total derivative writes
\begin{eqnarray}
\frac{D}{Dt}  =   (\vec{v}  \cdot   \vec{\nabla})  =  \frac{v_z}{\varpi_0}
\frac{d}{d \chi}
\end{eqnarray}
where $v_z$  is the  vertical velocity and  $\chi= z/\varpi_0$  is the
self-similar  variable that measures  the position  along a  flow line
anchored at $\varpi_0$.  With this in hand, and  the mass conservation
condition for an atomic  wind (Eq.~\ref{eq:mass}), the energy equation
Eq.~\ref{eq:en_cons} becomes an ODE along the flow line:
\begin{eqnarray} 
  \frac{d  \ln   T}{d \chi}  &=& \frac{2}{3} \frac{d\ln\tilde{n}}{ d \chi}  
  - \frac{d}{d \chi}\ln (1+ f_{\rm e}) + \frac{\Gamma  - \Lambda}{\frac{3}{2} n k
    T \frac{v_z}{\varpi_0}} \nonumber \\ 
  &\simeq &  \frac{2}{3} \frac{d \ln \tilde n}{d\chi} \left ( 1 - \frac{
      \Gamma_{\text{MHD}} + (\Gamma -\Lambda)_{\text{Map}}} 
    {\Lambda_{\text{adia}}} \right ) 
  \label{eq:energy}
\end{eqnarray}
The  term $d\ln(1+f_{\rm  e})/d\chi$  (due to  variations in  internal
energy density) is in fact negligible and has not been implemented for
computational  simplicity   (see  Appendix~\ref{ss:ign_h}).  The  term
$(\Gamma     -      \Lambda)_{\text{Map}}=     \Gamma_{\text{P}}     -
\Lambda_{\text{\rm      rad}}       -      \Lambda_{\text{col}}      -
\Lambda_{\text{rec}}$  is  provided by  {\tt  Mappings  Ic}. The  wind
thermal structure  is computed  by integrating along  a flow  line the
energy  equation~(\ref{eq:energy}) coupled  to the  set  of electronic
population equations (Eqs.~(\ref{eq:rate})~to (\ref{eq:el_cons})).

\subsubsection{Initial Conditions}

The   integration   of    the   set   equations   (\ref{eq:rate})   to
(\ref{eq:el_cons}) and (\ref{eq:energy}) along  the flow is an initial
value problem. Thus, some way  to estimate the initial temperature and
populations  must   be  devised.    All  calculations  start   at  the
slow-magnetosonic (SM)  point, which is  roughly at two  scale heights
above the disk midplane (for the solutions used here).

To  estimate the  initial temperature,  \citeauthor{Safier93a} equated
the poloidal flow  speed at the SM point to  the sound speed. Although
this estimation agrees with cold  flow theory, it is inconsistent with
the energy equation which is used further up in the jet.  Our approach
was  then to  compute the  initial temperature  and  ionic populations
assuming that
\begin{eqnarray}
\frac{D T}{Dt}\bigg\vert_{\chi=\chi_s}=0
&\ \ \ \mbox{and}\ \ \ & \frac{D  n_{A^i}}{Dt}\bigg\vert_{\chi=\chi_s} =0
\end{eqnarray}
is fulfilled at $\chi=\chi_s$. We thus assume for convenience that, at
the  base  of  the jet,  there  is  no  strong variations  neither  in
temperature nor  in ionization  fractions. Physically this  means that
the  gas is in  ionization equilibrium,  at the  obtained temperature,
with the  incoming radiation field.  The temperature  thus obtained is
always smaller than that provided by  the SM speed. This is due to the
large opening of the magnetic surfaces, providing a dominant adiabatic
cooling over all heating  processes. For numerical reasons the minimum
possible initial temperature  was set to 50~K.  It  is noteworthy that
the exact  value of the initial  temperature only affects  the base of
the     flow,     below     a     few    thousand     degrees     (see
Appendix~\ref{sss:t_init}). These  regions are too  cold to contribute
significantly   to  optical   line  emission,   leaving  observational
predictions unaffected.

The  initial populations are  computed by  {\tt Mappings  Ic} assuming
ionization equilibrium with the incoming radiation field.  However for
high accretion rates and for the outer zones of the wind, dust opacity
and  inclination  effects  shield  completely the  ionizing  radiation
field. The  temperature is  too low for  collisional ionization  to be
effective.   The  ionization  fraction  thus  reaches  our  prescribed
minimum --  all Na is in  the form \ion{Na}{ii} (Table~2)  and all the
other elements (computed by {\tt Mappings Ic}) neutral.  However, soon
the gas  flow gains height and  the ionization field  is strong enough
such  that  the  ionization  is  self-consistently  computed  by  {\tt
Mappings Ic}.

\subsubsection{Integration procedure}

After obtaining  an initial temperature  and ionization state  for the
gas we proceed by integrating the system of equations. In practice the
ionization evolution  is computed by {\tt Mappings  Ic} and separately
we integrate  Eq.~(\ref{eq:energy}) with a  Runge-Kutta type algorithm
(Press  et al.,  1988).  We  maintain  both the  populations and  {\tt
Mappings Ic} cooling/heating rates per $\tilde{n}^2$ fixed during each
temperature integrating spatial step.  After we call {\tt Mappings Ic}
to evolve  the populations and  rates, at the new  temperature, during
the time  taken by the fluid to  move the spatial step.   This step is
such, that  the RK integration  has a numerical accuracy  of $10^{-6}$
and, that the newly computed  temperature varies by less than a factor
of  $10^{-4}$.  Such  a small  variation in  temperature allows  us to
assume  constant  rates  and  populations  while  solving  the  energy
equation.  We checked  a few integrations by redoing  them at half the
step used and found that the error in the ionic fraction is $<10^{-3}$
in the jet, and $< 10^{-2}$ in the recollimation zone; the temperature
precision being roughly a few times better.  This ensures an intrinsic
numerical precision comfortably below  the accuracy of the atomic data
and the  $\mean{\sigma_{\rm{H}\,{\rm H}^+}v}$ collision cross-sections
which,  coupled to  abundance  incertitudes, are  the main  limitating
factors.   Details on  the  actual numerical  procedure  used by  {\tt
Mappings Ic} to compute the non-equilibrium gas evolution are given in
\cite{Binette85}.

%ch4
\section{Thermal structure results}\label{s:results}

In  this section  we  present the  calculated  thermal and  ionization
structure along  wind flow lines,  discuss the physical origin  of the
temperature  plateau  and  its  connection  with  the  underlying  MHD
solution,  discuss the  effect  of various  key  model parameters  and
finally    compare     our    results    with     those    found    by
\citeauthor{Safier93a}. The parameters  spanned for the calculation of
the thermal solutions are the wind ejection index $\xi$ describing the
flow line  geometry, the mass  accretion rate $\dot{M}_{\rm  acc}$ and
the cylindrical radius  $\varpi_0$ where the field is  anchored in the
disk.

\subsection{Temperature evolution}

\begin{figure}
   \resizebox{8.5cm}{!}{\includegraphics{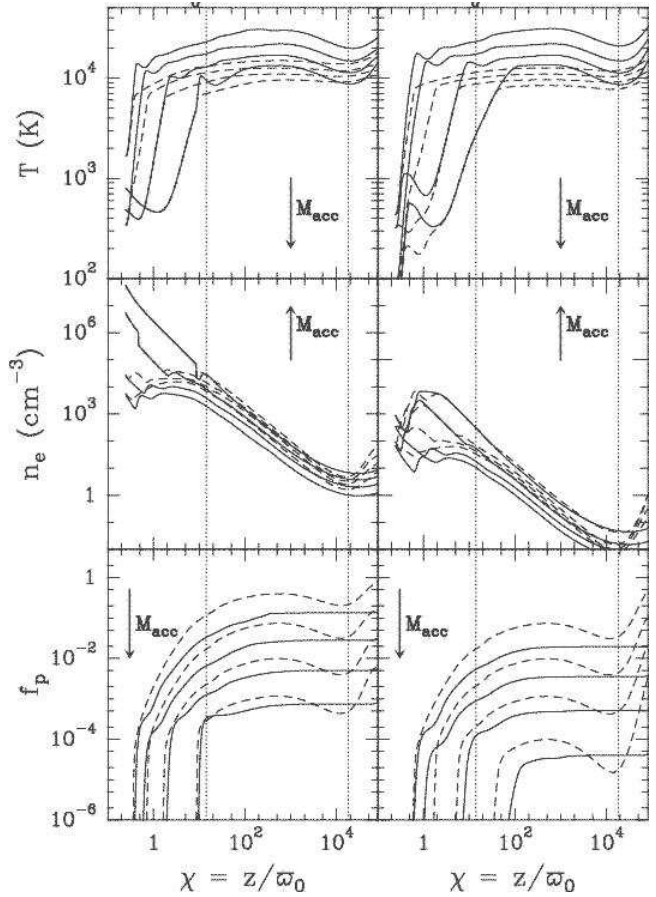}}
\caption{Several wind quantities versus  $\chi=z/\varpi_0$ for model B.
The out  of ionization equilibrium  calculations are the  solid curves
and, for  comparison, the ionization equilibrium are  the dashed ones.
The vertical  dotted lines mark  the Alfv\'en point  and recollimation
point.   {\bf  Top:}  Temperature,  {\bf Middle:}  electronic  density
$n_{\rm  e}$,  {\bf  Bottom:}   proton  fraction  $f_{\rm  p}=  n({\rm
H}^+)/n_{\rm H}$.  The accretion rate $\dot{M}_{\rm acc}$ increases in
the  direction of  the  arrow from  $10^{-8}$ to  $10^{-5}$M$_{\odot}$
yr$^{-1}$ in factors of 10.}
\label{fig:timion} %
\end{figure}

In Figure~\ref{fig:timion}, solid curves present the out of ionization
equilibrium evolution  of temperature, electronic  density, and proton
fraction  along flow  lines  with $\varpi_0  =  0.1$ and  1  AU, as  a
function  of $\chi  = z/\varpi_0$,  for accretion  rates  ranging from
$10^{-8}$  to  $10^{-5}{\rm   M}_{\odot}$  yr$^{-1}$.  For  comparison
purposes, dashed  curves plot the same  quantities calculated assuming
ionization equilibrium  at the local temperature  and radiation field.
For compactness we  present only these detailed results  for our model
B, with  an intermediate ejection index  $\xi = 0.007$.  We divide the
flow in three  regions: the base, the jet  and the recollimation zone.
These   regions  are  separated   by  the   Alfv\'en  point   and  the
recollimation point  (where the  axial distance reaches  its maximum).
We  only present  the initial  part  of the  recollimation zone  here,
because the dynamical solution is less reliable further out, where gas
pressure  is  increased  by  compression  and may  not  be  negligible
anymore.  Note  that the recollimation  zone was not yet  reached over
the    scales    of    interest    in   the    solutions    used    by
\citeauthor{Safier93a}.

%fig1______________________________________________________________
%_____________________________________________________________________

The  gas temperature  increases steeply  at  the wind  base (after  an
initial  cooling phase  for  high $\dot{M}_{\rm  acc} \ge  10^{-6}{\rm
M}_{\odot}$  yr$^{-1}$).  It  then  stabilizes in  a  hot  temperature
plateau around $T  \simeq 1-3 \times 10^4 K$,  before increasing again
after the recollimation point through compressive heating. The plateau
is  reached  further  out   for  larger  accretion  rates  and  larger
$\varpi_0$.  Its  temperature decreases with  increasing $\dot{M}_{\rm
acc}$.  The temperature  plateau and  its behavior  with $\dot{M}_{\rm
acc}$  were first  identified  by \citeauthor{Safier93a}  in his  wind
solutions.  We  will  discuss  in  Section~\ref{ss:plateau}  why  they
represent a  robust property of magnetically-driven  disk winds heated
by ambipolar diffusion.

\subsection{Ionization and electronic density}

The bottom  panels of  Fig.~\ref{fig:timion} plot the  proton fraction
$f_{\text p}= n({\rm H}^+)/n_{\rm H}$  along the flow lines.  It rises
steeply with wind temperature through collisional ionization, reaching
a value $\simeq 10^{-4}$ at  the beginning of the temperature plateau.
Beyond  this point,  it  continues  to increase  but  starts to  ``lag
behind'' the ionization  equilibrium calculations (dashed curves): the
density  decline in the  expanding wind  increases the  ionization and
recombination timescales.  Eventually, for $\chi \gtrsim 100$, density
is so low that these timescales become longer than the dynamical ones,
and the proton fraction becomes completely ``frozen-in'' at a constant
level,  typically a  factor 2-3  below the  value found  in ionization
equilibrium calculations (dashed curves).

The electron  density ($n_{\rm e}$)  evolution is shown in  the middle
panels of Fig.~\ref{fig:timion}.  In the jet region, where $f_{\rm p}$
is roughly  constant, the dominant  decreasing pattern with  $\chi$ is
set  by  the  wind  density   evolution  as  the  gas  speeds  up  and
expands. Similarly, the rise in  $n_{\rm e}$ in the recollimation zone
is due  to gas compression.  A remarkable result  is that, as  long as
ionization  is  dominated  by  hydrogen (i.e.   $f_{\text{p}}  \gtrsim
10^{-4}$), $n_{\rm e}$ is not highly dependent of $\dot{M}_{\rm acc}$,
increasing by  a factor of 10  only over three orders  in magnitude in
accretion rate.   This indicates a roughly inverse  scaling of $f_{\rm
p}$ with $\dot{M}_{\rm acc}$ (bottom panels of Fig.~\ref{fig:timion}),
a  property  already found  by  \citeauthor{Safier93a}  which we  will
discuss in more detail later.

\begin{figure}
  \resizebox{8.5cm}{!}{\includegraphics{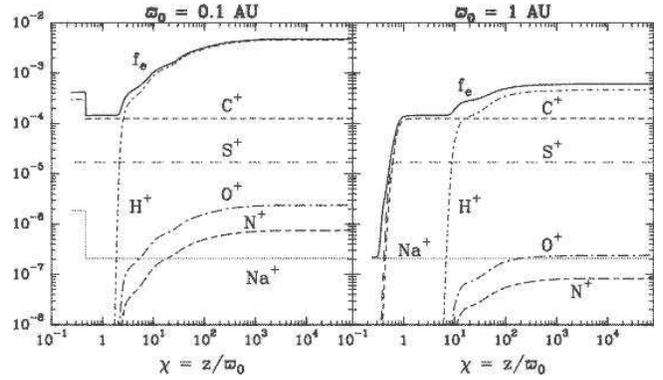}}
  \caption{Ion abundances with respect to hydrogen 
    ($f_{A^i}=n_{A^i}/\tilde{n}$  and thus  $f_{A^i}$ depends  also on
    the  abundances)   along  the   flow  line  versus   $\chi  \equiv
    z/\varpi_0$  for model B  in out  of ionization  equilibrium, with
    $\dot{M}_{\rm  acc}=10^{-6}  {\rm  M}_\odot \text{yr}^{-1}$.   The
    jump at  $\chi\sim0.5$ is due to  depletion as the  gas enters the
    sublimation surface.}
  \label{fig:timion_pop} %
\end{figure}

In regions at the wind  base where $f_{\text p} < 10^{-4}$, variations
of $n_{\rm  e}$ are  linked to the  detailed photoionization  of heavy
elements which are then  the dominant electron donors.  The respective
contributions  of  various  ionized  heavy  atoms  to  the  electronic
fraction $f_{\rm  e}$ is illustrated  in Fig.~\ref{fig:timion_pop} for
$\dot{M}_{\rm    acc}=10^{-6}{\rm   M}_{\odot}$    yr$^{-1}$.    While
\ion{O}{ii}   and  \ion{N}{ii}  are   strongly  coupled   to  hydrogen
collisional  ionization through charge  exchange reactions,  the other
elements are dominated by photoionization.  The sharp discontinuity in
\ion{C}{ii} and \ion{Na}{ii} at the wind base for $\varpi_0 = 0.1 $~AU
is  caused by  the crossing  of the  dust sublimation  surface  by the
streamline (see Appendix~\ref{ap:dust}).  Inside the surface we are in
the  dust sublimation  zone  where heavy  atoms  are consequently  not
depleted onto grains and hence  have a higher abundance.  In contrast,
for  $\varpi_0  =  1  $~AU,   the  flow  starts  already  outside  the
sublimation radius, in a region  well-shielded from the UV flux of the
boundary-layer,  where only Na  is ionized.   Extinction progressively
decreases as material is lifted  above the disk plane and sulfur, then
carbon, also become completely photoionized.

%fig2_______________________________________________________
%_____________________________________________________________________

\subsection{Heating and cooling processes}

\begin{figure*}
\begin{center}
   \resizebox{12cm}{!}{\includegraphics{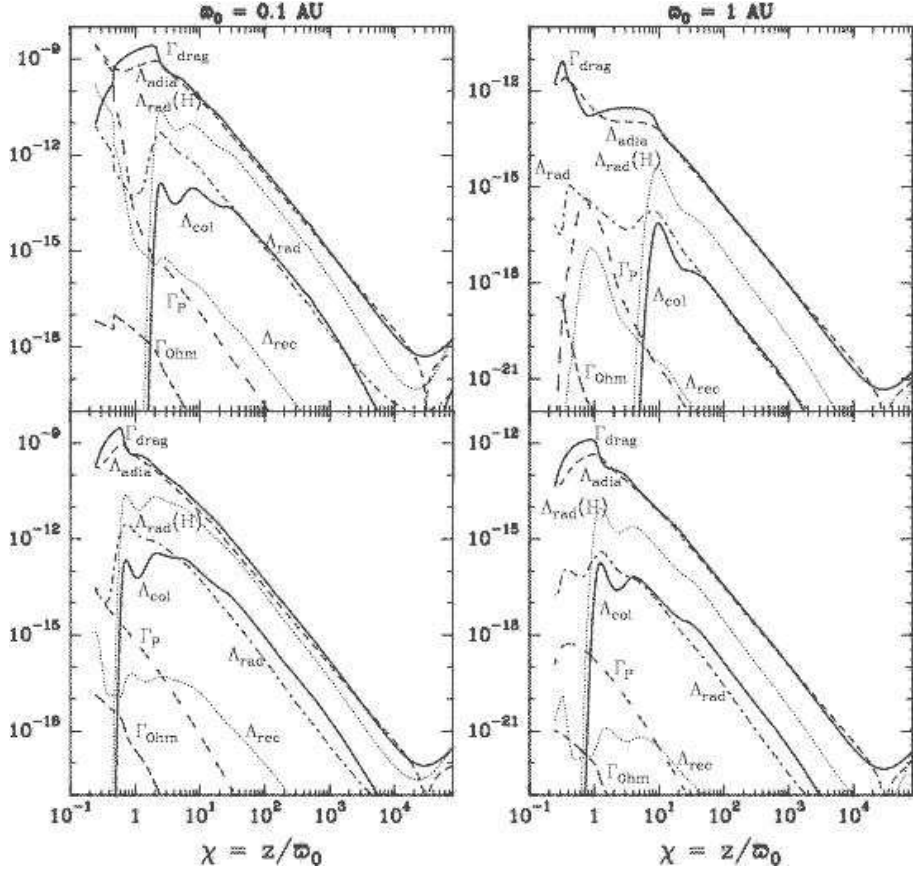}}
\caption{ Heating  and cooling  processes (in erg  s$^{-1}$ cm$^{-3}$)
along the flow line versus  $\chi \equiv z/\varpi_0$ for model B.  Top
figures for  $\dot{M}_{\rm acc}=10^{-6} {\rm  M}_\odot \text{yr}^{-1}$
and   bottom  ones  for   $\dot{M}_{\rm  acc}=10^{-7}   {\rm  M}_\odot
\text{yr}^{-1}$. Ambipolar heating and  adiabatic cooling appear to be
the dominant terms, although Hydrogen line cooling cannot be neglected
for the inner streamlines.}
\end{center}
\label{fig:timion_HC} %
\end{figure*}

The heating  and cooling  terms along the  streamlines for our  out of
equilibrium calculations  are plotted in  Fig.~\ref{fig:timion_HC} for
$\varpi_0 = 0.1 $ and 1  AU, and for two values of $\dot{M}_{\rm acc}$
= $10^{-6}$ and $10^{-7}$ M$_{\odot}$ yr$^{-1}$.

%fig3_______________________________________________________
%_____________________________________________________________________

Before the  recollimation point,  the main cooling  process throughout
the flow is adiabatic  cooling $\Lambda_{\rm adia}$, although Hydrogen
line   cooling  $\Lambda_{\rm   rad}({\rm  H})$   is   definitely  not
negligible.    The  main  heating   process  is   ambipolar  diffusion
$\Gamma_{\rm drag}$.  The  only exception occurs at the  wind base for
small  $\varpi_0  \le$  0.1   AU  and  large  $\dot{M}_{\rm  acc}  \ge
10^{-6}{\rm  M}_{\odot}$   yr$^{-1}$,  where  photoionization  heating
$\Gamma_{\rm   P}$  initially   dominates.   Under   such  conditions,
ambipolar diffusion heating is low  due to the high ion density, which
couples them  to neutrals and  reduces the drift responsible  for drag
heating.   However,  $\Gamma_{\rm P}$  decays  very  fast  due to  the
combined  effects of  radiation dilution,  dust opacity,  depletion of
heavy atoms in the dust phase, and the decrease in gas density. At the
same time,  the latter two  effects make $\Gamma_{\rm drag}$  rise and
become quickly  the dominant heating term. In  the recollimation zone,
the  main  cooling  process  is Hydrogen  line  cooling  $\Lambda_{\rm
rad}(\rm  H)$,  and  the  main  heating term  is  compression  heating
($\Lambda_{\rm adia}$ is negative).

A   striking  result  in   Fig.~\ref{fig:timion_HC},  also   found  by
\citeauthor{Safier93a}, is  that a close match  is quickly established
along  each streamline between  $\Lambda_{\rm adia}$  and $\Gamma_{\rm
drag}$, and is maintained until the recollimation region. The value of
$\chi$ where this balance is established is also where the temperature
plateau  starts.  We will  demonstrate below  why this  is so  for the
class of MHD wind solutions considered here.

\subsection{Physical origin of the temperature plateau}\label{ss:plateau}

%fig4_____________________________________________________________________ 
\begin{figure*}
\begin{center}
   \resizebox{18cm}{!}{\rotatebox{-90}{\includegraphics{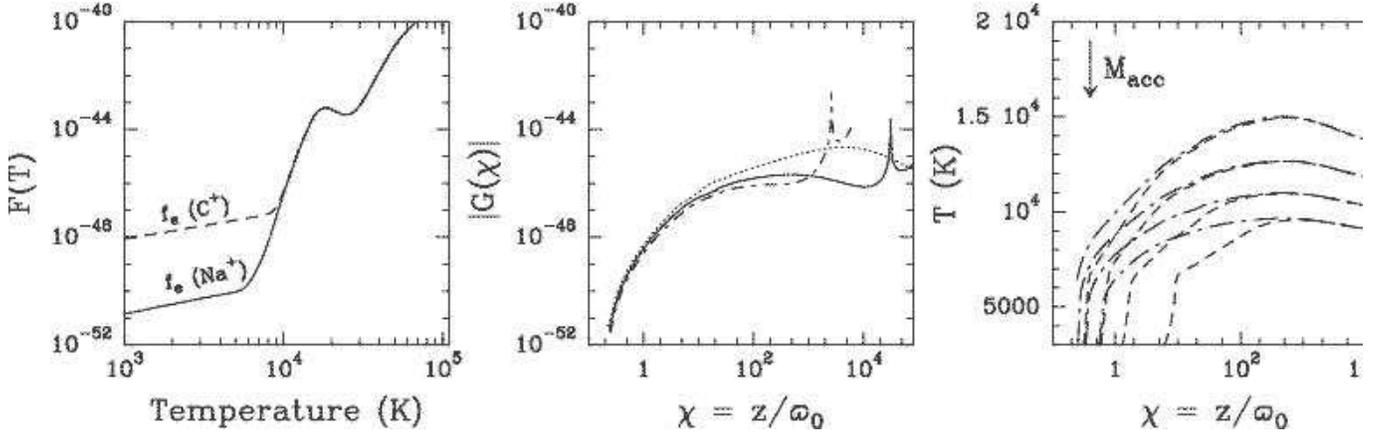}}}
\end{center}
\caption{ {\bf Left:}  Function $F(T)$ in erg~g~cm$^3$~s$^{-1}$
versus temperature {\em assuming  local ionization equilibrium} and an
ionization flux  that ionizes  only all Na  and all C.   {\bf Center:}
Function $G(\chi)$ in erg~g~cm$^3$~s$^{-1}$ for models A, B, C (bottom
to  top),  $\dot{M}_{\rm   acc}=10^{-6}  M_{\odot}{\rm  yr}^{-1}$  and
$\varpi_0=0.1{\rm AU}$. {\bf Right:}  Temperature for model B from the
complete calculations in ionization equilibrium (dashed), and assuming
$T   =T_{\ominus}$  as  given   by  eq.~\ref{eq:Tc}   (  dash-dotted).
$\varpi_0=0.1$  AU   and  accretion  rates   $\dot{M}_{\rm  acc}$  are
$10^{-8}$ to $10^{-5}$ M$_{\odot}$ yr$^{-1}$, from top to bottom.}
\label{fig:plateau}%
\end{figure*}
%_____________________________________________________________________ 
%

The existence of a  hot temperature plateau where $\Lambda_{\rm adia}$
exactly balances $\Gamma_{\rm drag}$ is the most remarkable and robust
property  of  magnetically-driven   disk  winds  heated  by  ambipolar
diffusion.   Furthermore, it occurs  throughout several  decades along
the flow including  the zone of the jet  that current observations are
able to spatially resolve.

In this section, we explore  in detail which generic properties of our
MHD  solution allow  a temperature  plateau at  $\simeq 10^4$~K  to be
reached, and  why this  equilibrium may not  be reached for  other MHD
wind solutions.

\subsubsection{Context}

First, we  note that the energy equation  (Eq.~\ref{eq:energy}) in the
region where drag heating and adiabatic cooling are the dominant terms
(which  includes the  plateau region)  can be  cast in  the simplified
form:
\begin{eqnarray}
\frac{d  \ln  T}{d \ln \chi} &=&
-\frac{2}{3}  \frac{ d  \ln  \tilde{n}}{ d \ln \chi} \times
\Big(\frac{\Gamma_{\rm drag}}{\Lambda_{\rm adia}}-1\Big) \nonumber \\
&=& \delta^{-1} \times
\Big(\frac{G(\chi)}{F(T)}-1\Big), \label{eq:amb_temp}
\end{eqnarray}
where:
\begin{eqnarray}\label{eq:delta}
\delta(\chi)^{-1}       \equiv       \Big(-\frac{2}{3}\frac{d      \ln
  \tilde{n}}{d\ln\chi}\Big),
\end{eqnarray}
remains  positive before  recollimation  and depends  only  on the  MHD
solution,
\begin{eqnarray}\label{eq:Gdef}
G(\chi)=-\frac{\frac{1}{c^2}  \bigl\| \vec{J}  \times  \vec{B} \bigr\|^2}
{\tilde{n}^2        (\vec{v}\cdot\vec{\nabla})\tilde{n}}       \propto
\frac{M_{\star}^2}{\dot{M}_{\text{acc}} \varpi_0}
\end{eqnarray}
is a positive function, before recollimation, that depends only on the
MHD wind solution, and
\begin{eqnarray}
F(T)={kT(1+f_{\rm e})}{\overline{m_{in} f_i f_n
\mean{\sigma_{in}v}}}
\times {\Big( \frac{ \rho }{ \overline{\rho}_n } \Big)^2}
\end{eqnarray}
also positive,  depends only on  the local temperature  and ionization
state  of   the  gas.   The   functions  $G$  and  $F$   separate  the
contributions  of the  MHD dynamics  and ionization  processes  in the
final thermal solution.

The  function $\delta$  is  roughly constant  and  around unity  before
recollimation,  it diverges  at  the recollimation  point and  becomes
negative after it. Throughout the plateau $\delta \sim 1$.

The  ``wind  function''  $G$  is   plotted  in  the  center  panel  of
Fig.~\ref{fig:plateau} for our  3 solutions.  It rises by  5 orders of
magnitude  at the  wind base  and then  stabilizes in  the  jet region
(until  it diverges  to infinity  near the  recollimation  point). The
physical reason  for its behavior is  better seen if we  note that the
main force driving the flow is the Lorentz force:
\begin{equation}
G\propto \Big|\Big|\frac{D\vec{v}}{Dt}\Big|\Big|^2 \times 
\Big(\frac{D\rho}{Dt}\Big)^{-1}.
\end{equation}
For  an  expanding  and  accelerating  flow  $(D\rho/Dt)^{-1}$  is  an
increasing function.   At the wind base the  Lorentz force accelerates
the gas thus causing a fast  increase in $G$.  Once the Alfv\'en point
is   reached,   the   acceleration   is  smaller   and   $D\vec{v}/Dt$
decreases. However this decrease is not  so abrupt as in the case of a
spherical  wind, because  the Lorentz  force  is still  at work,  both
accelerating  and  collimating  the  flow. This  collimation  in  turn
reduces the rate of  increase of $(D\rho/Dt)^{-1}$.  The stabilization
of $G$ observed after the Alfv\'en point is thus closely linked to the
jet dynamics.

The ``ionization function'' $F$ is in general a rising function of $T$
and is plotted  in the left panel of  Fig.~\ref{fig:plateau} under the
approximation  of  local  ionization  equilibrium.   Two  regimes  are
present: In the low temperature regime, $f_{\rm i}$~$\gg f_{\rm p}$ is
dominated by  the abundance of  photoionized heavy elements  and $F(T)
\propto T\,f_{\rm  i}$ increases linearly with $T$,  for fixed $f_{\rm
i}$.  The effect of the UV  flux in this region is to shift vertically
$F(T)$: for a low UV flux regime only Na is ionized and $ f_{\text{i}}
\simeq  f(\ion{Na}{ii})$; for  a high  UV flux  regime were  Carbon is
fully  ionized,  $f_{\text{i}}  \simeq  f(\ion{C}{ii})$. In  the  high
temperature  regime  ($T  \geq  8000$~K)  where  hydrogen  collisional
ionization  dominates,  $f_{\rm  i}$  $\simeq f_{\rm  p}$,  and  $F(T)
\propto   T\,f_{\rm  p}$   becomes  a   steeply  rising   function  of
temperature, until hydrogen is  fully ionized around $T \simeq 2\times
10^4$~K.   The  following second  rise  in  $F(T)$  is due  to  Helium
collisional  ionization. As  we  go out  of  perfect local  ionization
equilibrium the effect is to decrease  the slope of F(T) in the region
where H ionization dominates.   In the extreme situation of ionization
freezing, $F(T)$  becomes linear again as in  the photoionized region:
$F(T) \propto T f_{\text{p,freezed}}$.

\subsubsection{Conditions needed for a hot plateau}

The plateau  is simply  a region where  the temperature does  not vary
much,
\begin{eqnarray}\label{eq:p_def}
\frac{d \ln T}{d \ln \chi} = \epsilon 
{\hspace{1cm }\rm with \hspace{1cm }} |\epsilon|\ll 1.
\end{eqnarray}
Naively, temperatures $T_\ominus(\chi)$ defined by,
\begin{eqnarray}\label{eq:Tc}
F(T_\ominus)= G(\chi) &\Longleftrightarrow &\Gamma_{\rm{drag}}
= \Lambda_{\rm{adia}},
\end{eqnarray}
will  zero  the right  hand  side  of  Eq.~\ref{eq:amb_temp} and  thus
satisfy  the  plateau condition.   This  equality  is  the {\em  first
constraint}  on the  wind functiond  $G$, because  there must  exist a
temperature $T_\ominus$  such that  the equality holds.   However this
condition is  not sufficient.  Indeed  the above equality  describes a
curve\footnote{This is  only true if  $F$ is a monotonous  function of
$T$.  As  can be seen  from Figure~\ref{fig:plateau} this is  true for
almost all its domain.}  $T_\ominus(\chi)$ which must be flat in order
to satisfy the  plateau condition (Eq.~\ref{eq:p_def}).  Therefore the
requirement  of  a  flat  $T_\ominus$  translates  in  a  {\em  second
constraint}  on  the variation  of  $G$  with  respect to  $F$.   This
constraint is obtained by differentiating Eq.~\ref{eq:Tc}. We obtain
\begin{eqnarray}\label{eq:p_dif}
\frac{d \ln  G}{d \ln \chi} =  \frac{d \ln F}{d  \ln
  T}\bigg\vert_{T = T_\ominus} \times \epsilon
\end{eqnarray}
and after using (Eq.~\ref{eq:p_def}):
\begin{eqnarray}
\|\frac{d \ln G}{d \ln \chi}\| \ll \|\frac{d \ln F}{d \ln T} \bigg\vert_{T
  = T_\ominus}(\chi)\|. 
\end{eqnarray}
{\em Thus  only winds where the  wind function $G$  varies much slower
than the  ionization function  $F$ will produce  a plateau.}   This is
fulfilled for  our models:  Below the Alfv\'en  surface, $G$  varies a
lot, but collisional H  ionization is sufficiently close to ionization
equilibrium   that  $F(T)$   still  rises   steeply   around  $10^4$~K
(Fig.~\ref{fig:plateau}).  For our numerical values of $G$, within our
range  of $\dot{M}_{\text{acc}}$  and $\varpi_0$,  we  have $T_\ominus
\simeq 10^4$~K and thus $|d \ln G/d  \ln \chi| \ll |d \ln F/d \ln T|$.
Further out, where ionization is frozen  out, we have $d \ln F/d \ln T
= 1$ (because  $F\propto T f_{\rm p, freezed}$) but  it turns out that
in this region $G$ is a slowly varying function of $\chi$, and thus we
still have $|d \ln G/d \ln \chi| \ll |d \ln F/d \ln T|$.

Finally, a {\it third constraint}  is that the flow must quickly reach
the plateau solution $T(\chi)  = T_\ominus(\chi)$ and tend to maintain
this equilibrium.   Let us assume  that $T=T_\ominus$ is  fulfilled at
$\chi=\chi_0$, what  will be the temperature  at $\chi= \chi_0(1+x)$~?
Letting $T=  T_\ominus(1 + \vartheta)$ and assuming  $\epsilon \ll 1$,
Eq.~\ref{eq:amb_temp} gives  us $\vartheta = \exp(-  \alpha x)$, which
provides an exponential convergence towards $T=T_\ominus$ as long as
\begin{eqnarray}
\alpha= \frac{1}{\delta} \frac{d \ln  F}{d \ln T}\bigg\vert_{T = T_\ominus}
> 0 \ . 
\end{eqnarray}
Note that $\alpha$ depends mainly  on the MHD solution (independent of
$\dot{M}_{\rm acc}$/$\varpi_0$) and that the steeper the function $F$,
the faster  the convergence.  The above criterion  is always fulfilled
in the expanding region of  our atomic wind solutions, where $\delta >
0$ and  $F$ increases with  temperature.  The physical reason  for the
convergence can be  easily understood in the following  way: it can be
readily  seen that  if at  a  given point  $T> T_\ominus(\chi)$,  then
$G(\chi)/F(T) <  1$ and the  gas will cool  (cf. Eq.~\ref{eq:amb_temp}
with $\delta > 0$).   Conversely, if $T<T_\ominus(\chi)$, the gas will
heat up.   Thus, for $\delta  > 0$, the  fact that $F(T)$ is  a rising
function  introduces   a  feedback  that  brings   and  maintains  the
temperature  close to its  local equilibrium  value $T_\ominus(\chi)$,
and $\Lambda_{\rm adia}$ close to $\Gamma_{\rm drag}$.

We conclude that three analytical criteria must be met by any MHD wind
dominated  by ambipolar  diffusion heating  and adiabatic  cooling, in
order to converge to a hot temperature plateau:\\
(1) Equilibrium:   the   wind  function   $G$   must   be  such   that
$F(T)=G(\chi)$ is possible around $T \simeq 10^4$~K; \\
(2) Small temperature variation: the wind function $G(\chi)$ must vary
slower than the ionization function $F(T)$ such that \mbox{$|{ d \ln G
/d\ln \chi}| \ll |d\ln F/d\ln T|$}: (i) the wind must be in ionization
equilibrium, or  near it, in regions  where $G$ is a  fast function of
$\chi$; (ii) once  we have ionization freezing, $G$  must vary slowly,
with \mbox{$| d \ln G / d\ln \chi | \ll 1$};\\
(3) Convergence: $\alpha >0$,  i.e. ${2 \over 3} (d  \ln \tilde{n} / d
\ln \chi) \times (d \ln F / d \ln T) \vert_{T = T_\ominus} < 0$, which
is always verified for an atomic and expanding wind.

\subsubsection{Comments on other MHD winds}

Not all types  of MHD wind solutions will  verify our first criterion.
Physically, the  large values of  $G(\chi)$ observed in  our solutions
indicates that there is still a non-negligible Lorentz force after the
Alfv\'en surface. In  this region (which we call  the jet) the Lorentz
force is dominated by its poloidal component which both collimates and
accelerates  the  gas  (Fig.~\ref{fig:adim}).   The  gas  acceleration
translates in a further decrease  in density contributing to a further
increase in  $G$.  Models that  provide most of the  flow acceleration
before  the Alfv\'en  surface  might turn  out  to have  a lower  wind
function $G(\chi)$, not numerically  compatible with the steep portion
of the ionization function $F(T)$.  These models would not establish a
temperature   plateau   around   $10^4$~K   by   ambipolar   diffusion
heating. They  would either stabilize  on a lower  temperature plateau
(on the linear part of $F(T)$) if our second criterion is verified, or
continue to  cool if $G$ varies  too fast for the  second criterion to
hold.  This is  the case in particular for  the analytical wind models
considered  by \cite{Ruden90},  where the  drag force  was  computed a
posteriori  from a prescribed  velocity field.   The $G$  function for
their  parameter space  (Table~3  of  Ruden et  al.)   peaks at  $\sim
10^{-48}$~erg~g~cm$^3$~s$^{-1}$ at $\sim 3 R_{\star}$ and then rapidly
decreases  as $G \propto  r^{-1}$ for  higher radii.   This translates
into a cooling wind without a plateau.
 
\subsection{Scalings of plateau parameters with $\dot{M}_{\rm acc}$ and $\varpi_0$}

%fig5_____________________________________________________________________ 
\begin{figure} 
\begin{center}
   \resizebox{8cm}{!}{\rotatebox{-90}{\includegraphics{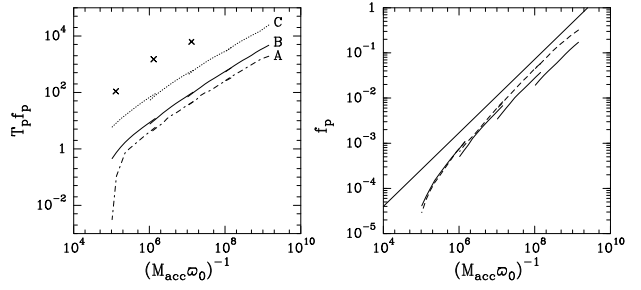}}}
\caption{
Verification  of  the  plateau scalings  (Eq.~\ref{eq:p_corr}).   {\bf
Left:}  we plot the  measured $T_\ominus  f_{\text{p},\ominus}$ versus
$(\dot{M}_{\text{acc}}\varpi_0)^{-1}$  for   all  models  in   out  of
ionization equilibrium.   The evolution is linear except  for the very
edges  of the  $(\dot{M}_{\text{acc}}\varpi_0)^{-1}$ domain.   This is
due to the failure of our  assumptions: in the lower edge $f_{\rm e} <
10^{-4}$ and thus  it isn't dominated by H; in  the upper edge $f_{\rm
e} > 0.1 $ and thus the small ionization fraction approximation fails.
The  crosses are  from  \citeauthor{Safier93a}; because  of a  smaller
momentum transfer rate coefficient they are quite above model C.  {\bf
Right:}   we   plot   the   measured   $f_{\text{p},\ominus}$   versus
$(\dot{M}_{\text{acc}}\varpi_0)^{-1}$ for Model B in out of ionization
equilibrium  (solid) and ionization  equilibrium (dashed).   Note that
all  the  variation is  absorbed  by  $f_{\text{p},\ominus}$ and  thus
$f_{\text{p},\ominus}  \propto (\dot{M}_{\text{acc}}\,\varpi_0)^{-1}$.
A straight line is also plot for comparison.}
\label{fig:p_corr} %
\end{center}
\end{figure}
%_____________________________________________________________________
%

The   balance    between   drag   heating    and   adiabatic   cooling
(Eq.~\ref{eq:Tc}) can  further be used  to understand the  scalings of
the  plateau temperature  $T_{\ominus}$ and  proton  fraction $f_{{\rm
p},\ominus}$ with the accretion rate $\dot{M}_{\rm acc}$ and flow line
footpoint radius  ($\varpi_0$).  In the plateau  region, ionization is
intermediate, i.e.,  sufficiently high to be dominated  by protons but
with most  of the  Hydrogen neutral.  Under  these conditions  we have
$F(T)\propto  T_{\ominus} f_{{\rm  p},\ominus}$.  On  the  other hand,
self-similar    disk   wind    models    display   $G(\chi)    \propto
(\dot{M}_{\text{acc}}\varpi_0)^{-1}$  (Eq.~\ref{eq:Gdef}).  Therefore,
we expect
\begin{eqnarray}\label{eq:p_corr}
T_\ominus     f_{\rm       p}        &\propto       &\frac{1}
{\dot{M}_{\text{acc}}\,\varpi_0}.
\end{eqnarray}
This behavior  is verified in the left  panel of Fig.~\ref{fig:p_corr}
for our out-of-equilibrium results for the 3 wind solutions.

To predict how much of  this scaling will be absorbed by $T_{\ominus}$
and  how  much   by  $  f_{{\rm  p},\ominus}$,  \citeauthor{Safier93a}
considered   the  ionization   equilibrium   approximation:  For   the
temperature  range of  the plateau  ($T\sim10^4$ K)  $f_{\rm p}\simeq$
$f_{\rm  i}$ is  a  very fast  varying  function of  $T$  that can  be
approximated as $f_{\rm p}\propto T^{a}$ with $a \gg 1$.  One predicts
that  $T_{\ominus}\propto (\dot{M}_{\text{acc}}\,\varpi_0)^{-1/(a+1)}$
while                    $f_{{\rm                   p},\ominus}\propto
(\dot{M}_{\text{acc}}\,\varpi_0)^{-a/(a+1)}                      \simeq
(\dot{M}_{\text{acc}}\,\varpi_0)^{-1}$.   Hence,  the inverse  scaling
with  ($\dot{M}_{\rm  acc}\varpi_0$)  should  be  mostly  absorbed  by
$f_{{\rm p},\ominus}$,  while the  plateau temperature is  only weakly
dependent on these parameters.  This is verified in the right panel of
Fig.~\ref{fig:p_corr},  where   $f_{{\rm  p},\ominus}$  in  ionization
equilibrium      is      plotted       as      a      function      of
$(\dot{M}_{\text{acc}}\,\varpi_0)^{-1}$.   The  predicted  scaling  is
indeed closely followed.

Let us  now turn back  to the actual  out-of-equilibrium calculations.
At  the base of  the flow  we find  that the  wind evolves  roughly in
ionization  equilibrium (see  Figure~\ref{fig:timion}),  however at  a
certain point the ionization fraction  freezes at values that are near
those of the ionization equilibrium zone. This effect implies that the
ionization fraction should roughly scale as the ionization equilibrium
values  {\em at  the upper  wind base}.   This in  indeed  observed in
Fig.~\ref{fig:p_corr}.   This  memory  of the  ionization  equilibrium
values   by  $f_{\rm   p}$  (as   observed  in   the  solar   wind  by
\cite{Owockietal83})  is  the  reason  why the  scalings  of  $f_{{\rm
p},\ominus}$  with ($\dot{M}_{\rm  acc}\varpi_0$) remain  correct.  We
computed  for  our solutions  the  scalings  and  found for  model  B:
$f_{\text{p},\ominus}\propto\dot{M}_{\text{acc}}^{-0.76}$,
$f_{\text{p},\ominus}\propto\varpi_0^{-0.83}$,
$T_{\ominus}\propto\dot{M}_{\text{acc}}^{-0.13}$ and  no dependence of
$T$  on $\varpi_0$,  confirming the  memory effect  on  the ionization
fraction only.

Finally, we  note that for  accretion rates in  excess of a  few times
$10^{-5}$ M$_\odot$  yr$^{-1}$, the hot plateau should  not be present
anymore: Because  of its inverse  scaling with $\dot{M}_{\text{acc}}$,
the  wind  function  $G$  remains  below  $10^{-47}$  erg  g  cm$^{3}$
s$^{-1}$,  and  $F   =  G$  occurs  below  $10^4$~K,   on  the  linear
low-temperature part of $F(T)$ where $f_{\rm i} \simeq f(\ion{C}{ii})$
(see Figure~\ref{fig:plateau}).  These  colder jets will presumably be
partly  molecular.   Interestingly,  molecular  jets  have  only  been
observed  so far  in  embedded protostars  with  high accretion  rates
(e.g. Gueth \& Guilloteau 1999).

\subsection{Effect of the ejection index $\xi$}

The  importance  of the  underlying  MHD  solution  is illustrated  in
figure~\ref{fig:plateau}. The ejection  index $\xi$ is directly linked
to the mass loaded in  the jet ($\kappa$, see Tab.~\ref{tab:param} and
Ferreira  1997).   Thus  a  higher  $\xi$ translates  in  an  stronger
adiabatic cooling  because more mass is being  ejected.  The ambipolar
diffusion heating is less sensitive to the ejection index, because the
density increase is balanced by a stronger magnetic field.  Hence, the
wind function $G(\chi)$ decreases with increasing $\xi$.  As a result,
the  plateau temperature  and ionization  fraction also  decrease (see
Eq.~\ref{eq:p_corr}).

\begin{figure}
\begin{center}
   \resizebox{8cm}{!}{\includegraphics{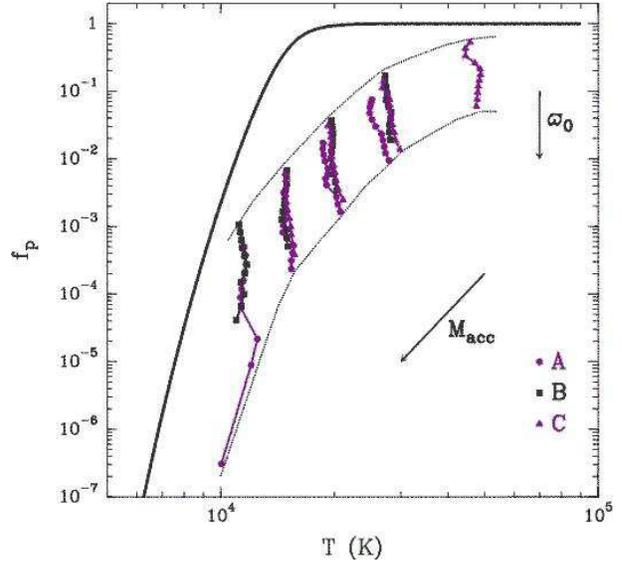}}
\caption{ Out  of ionization equilibrium evolution  of wind quantities
{\em   in   the   plateau}.    Points   are   for   all   flow   lines
($\varpi_0=0.07\times1.3^i$ AU and $i=0,1,...,10$) and accretion rates
($\dot{M}_{\rm   acc}=10^{-i}\:{\rm   M}_\odot{\text{yr}}^{-1}$   with
$i=5,6,7,8$).  We plot,  $f_{\text{p},\ominus}$ versus $T_\ominus$ for
all models C (red), B (black), A (blue) and in red the values expected
from ionization  equilibrium.  The dashed/dotted lines  are for models
without depletion, the solid lines are for models with depletion.  The
accretion rates  increase in the  direction top right to  bottom left.
The thick solid curve traces $f_{\text{p}}$ in ionization equilibrium,
while the two dotted lines embrace the locus of our MHD solutions.}
\label{fig:oie_plateau}%
\end{center}
\end{figure}

In figure~\ref{fig:oie_plateau} we summarize our results for the three
models   by  plotting   the  plateau   $f_{{\rm   p},\ominus}$  versus
$T_{\ominus}$, for  several $\dot{M}_{\rm acc}$  (values of $\varpi_0$
are connected  together).  In this plane,  our MHD solutions  lie in a
well-defined ``strip'' located below the ionization equilibrium curve,
between the  two dotted curves.   For a given model,  as $\dot{M}_{\rm
acc}$ increases,  the plateau ionization fraction  and the temperature
both decrease,  as expected from the scalings  discussed above, moving
the model  to the  lower-left of the  strip.  Increasing  the ejection
index decreases $G(\chi)$, and it can  be seen that this has a similar
effect as increasing $\dot{M}_{\rm acc}$ (Eq.~\ref{eq:p_corr}).

%fig7______________________________________________________________________ 
%
%______________________________________________________________________ 

\subsection{Depletion effects on the thermal structure}

In our  calculations we take  into account depletion of  heavy species
into the dust  phase. We ran our model with  and without depletion and
found  these  effects  to  be  minor.  Changes  are  only  found  when
$f_{\text{p}} \lesssim 10^{-4}$.  The temperature without depletion is
slightly   reduced    (the   higher   ionization    fraction   reduces
$\Gamma_{\text{drag}}$) and  as a consequence $f_{\text{p}}  $ is also
smaller.  Normally  these changes  affect only the  wind base,  as the
temperature increases  $f_{\text{p}}$ dominates the  ionization and we
obtain  the same  results  for  the plateau  zone.   However for  high
accretion    rates   ($\dot{M}_{\rm    acc}=10^{-5}    {\rm   M}_\odot
\text{yr}^{-1}$) in  the outer wind  zone (large $\varpi_0$)  we still
have  $f_{\text{p}} \lesssim  10^{-4}$ for  the plateau  and  thus the
temperature without depletion is reduced there.

\subsection{Differences with the results of \citeauthor{Safier93a}}\label{ss:dif_s}

%fig8______________________________________________________________________ 
\begin{figure}
\begin{center}
   \resizebox{8cm}{!}{\rotatebox{-90}{\includegraphics{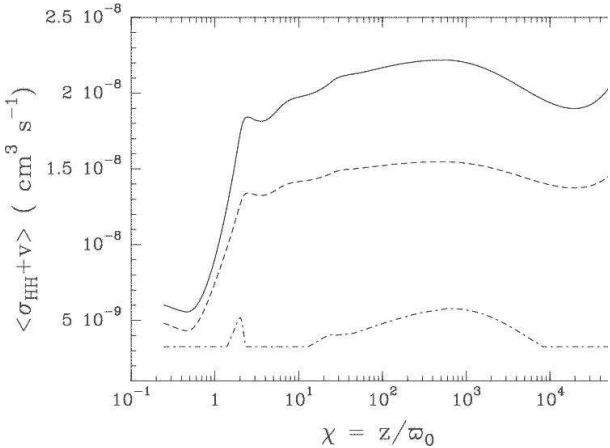}}}
\end{center}
\caption{$\mean{\sigma_{\mbox{\tiny  H   H$^+$}}v}$  in
cm$^{3}$  s$^{-1}$ using  Draine  expression (solid),  using Geiss  \&
Buergi expression  (see Appendix~\ref{ap:crosssec}) (dash)  and Draine
expression but ignoring the thermal velocity (dot).}
\label{fig:cross} %
\end{figure}
%______________________________________________________________________ 

The  most  striking  difference  between  our  results  and  those  of
\citeauthor{Safier93a}  is  an ionization  fraction  10  to 100  times
smaller.    This  difference   is   mainly  due   to  both   different
$\mean{\sigma_{\text{H H}^+}  v}$ momentum transfer  rate coefficients
and dynamical MHD wind models.

The  critical importance  of  the momentum  transfer rate  coefficient
($\mean{\sigma_{\text{H H}^+} v}$) for the plateau ionization fraction
can be  seen by  repeating the reasoning  in the previous  section but
including the  momentum transfer rate coefficient in  the scalings. We
thus obtain
\begin{eqnarray}
T_\ominus\,f_{\text{p},\ominus}                      \propto
\frac{1}{\mean{\sigma_{\text{H H}^+} v} \dot{M}_{\text{acc}} \varpi_0}
\ .
\end{eqnarray}
This shows  that, because the  freezing of the ionization  fraction is
correlated to the  ionization fraction at the base  of the wind (which
is  in ionization  equilibrium),  $f_{\text{p}}$ will  scale with  the
momentum  transfer rate  coefficient value.   This means  that  if the
momentum  transfer  rate coefficient  is  larger,  there  is a  better
coupling between ions  and neutrals and hence a  smaller drag heating.
For   the  calculation  of   the  $\mean{\sigma_{\text{H   H}^+}  v}$,
\citeauthor{Safier93a}  ignored   the  contribution  of   the  thermal
velocity  in  the collisional  relative  velocity.  This  considerably
reduces   $\mean{\sigma_{\text{H  H}^+}   v}$   and  thus,   increases
$f_{\text{p}}$.  In  figure~\ref{fig:cross} we plot  the corresponding
momentum  transfer  rate coefficient  values.   It  can  be seen  that
ignoring  the  thermal  contribution  to the  momentum  transfer  rate
coefficient decreases  it typically  by a factor  of $\gtrsim  6$.  We
also plot  in this figure  the value obtained  by \cite{GeissBuergi86}
illustrating   the  uncertainties  in   the  momentum   transfer  rate
coefficient (more on this in Appendix~\ref{ap:crosssec}).

\section{Concluding remarks}

We performed detailed non-equilibrium  calculations of the thermal and
ionization structure of  atomic, self-similar magnetically-driven jets
from  keplerian accretion  disks. Current  dissipation  in ion-neutral
collisions -- ambipolar diffusion--,  was assumed as the major heating
source.  Improvements  over  the  original work  of  \citep{Safier93a,
Safier93b} include: a)  detailed dynamical models by \cite{Ferreira97}
where  the  disk is  self-consistently  taken  into  account but  each
magnetic surface  assumed isothermal; b) ionization  evolution for all
relevant ``heavy  atoms'' using {\tt  Mappings Ic} code;  c) radiation
cooling by  hydrogen lines, recombination  and photoionization heating
using  {\tt Mappings  Ic}  code; d)  H-H$^+$  momentum exchange  rates
including   thermal   contributions;  and   e)   more  detailed   dust
description.

We obtain, as \citeauthor{Safier93a}, warm jets with a hot temperature
plateau at  $T\simeq 10^4$~K. Such a  plateau is a  robust property of
the atomic disk winds considered  here for accretion rates less than a
few  times  $10^{-5}  {\rm  M}_\odot$  yr $^{-1}$.   It  is  a  direct
consequence  of  the  characteristic  behavior of  the  wind  function
$G(\chi)$ defined in  Eq.~\ref{eq:Gdef}: (i) $G(\chi)$ increases first
and  becomes  larger  than  a  certain  value  fixed  by  the  minimum
ionization  fraction (see  Fig.~\ref{fig:plateau}); (ii)  $G(\chi)$ is
flat  whenever  ionization freezing  occurs  (collimated jet  region).
More generally,  we formulate three  analytical criteria that  must be
met  by any  MHD wind  dominated  by ambipolar  diffusion heating  and
adiabatic cooling in order to converge to a hot temperature plateau.

The scalings  of ionization fractions and temperatures  in the plateau
with     $\dot{M}_{\rm    acc}$     and     $\varpi_0$    found     by
\citeauthor{Safier93a}   are  recovered.    However   the  ionizations
fractions are 10 to 100  times smaller, due to larger H-H$^+$ momentum
exchange   rates  (which   include  the   dominant   thermal  velocity
contribution ignored  by \citeauthor{Safier93a}) and  to different MHD
wind dynamics.

We performed  detailed consistency checks for our  solutions and found
that  local  charge   neutrality,  gas  thermalization,  single  fluid
description  and ideal MHD  approximation are  always verified  by our
solutions. However at low accretion  rates, for the base of outer wind
regions ($\varpi_0\sim 1$AU) and increasingly for higher $\xi$, single
fluid calculations  become questionable.  For  the kind of  jets under
study,  a multi-component  description  is necessary  for field  lines
anchored  after $\varpi_0  >  1$  AU.  So  far,  all jet  calculations
assumed  either isothermal  or  adiabatic magnetic  surfaces. But  our
thermal computations  showed such an increase in  jet temperature that
thermal pressure  gradients might become relevant in  jet dynamics. We
therefore checked  the ``cold''  fluid approximation by  computing the
ratio of  the thermal  pressure gradient to  the Lorentz  force, along
($\beta_\parallel$) and perpendicular  ($\beta_{\perp}$) to a magnetic
surface. Both ratios increase for lower accretion rates and outer wind
regions. We found that  for some solutions, thermal pressure gradients
play  indeed  a  role,  however   only  at  the  wind  base  (possible
acceleration) and in the  recollimation zone (possible support against
recollimation).   Fortunately,  \citepalias[as   will  be  seen  in  a
companion  paper,][]{Garcia2001b}, the  dynamical solutions  which are
found inconsistent are also those rejected on an observational ground.
Therefore, it turns out that the models that best fit observations are
indeed consistent.

\begin{acknowledgements}

        PJVG acknowledges financial support from Funda\c{c}\~ao para a
        Ci\^encia  e Tecnologia by  the PRAXIS  XXI/BD/5780/95, PRAXIS
        XXI/BPD/20179/99 grants.  The work  of LB was supported by the
        CONACyT grant 32139-E. We  thank the referee, Pedro N. Safier,
        for his helpful  comments. We acknowledge fruitful discussions
        with  Alex  Raga,  Eliana   Pinho,  Pierre  Ferruit  and  Eric
        Thi\'ebaut.  We are grateful to Bruce Draine for communicating
        his grain  opacity data, and  to Pierre Ferruit  for providing
        his C interface to the {\tt Mappings Ic} routine.  PJVG warmly
        thanks the  Airi team and  his adviser, Renaud Foy,  for their
        constant support.

\end{acknowledgements}

\appendix
\section{Multicomponent MHD equations}\label{ap:dissipa}

\subsection{Single fluid description}\label{ap:heat}

Let  us consider  a fluid  composed of  $\alpha$ species  of numerical
density  $n_\alpha$, mass $m_\alpha$,  charge $q_\alpha$  and velocity
$\vec v_\alpha$.  All species are assumed to be coupled enough so that
they have the same temperature $T$. To get a single fluid description,
we then define
\begin{eqnarray}
\rho &=& \Sigma_\alpha n_\alpha m_\alpha \nonumber \\
\rho \vec v &= & \Sigma_\alpha m_\alpha n_\alpha \vec v_\alpha \nonumber\\
P &= &  \Sigma_\alpha n_\alpha k_{\text{B}} T \\
\vec J &=& \Sigma_\alpha n_\alpha q_\alpha \vec v_\alpha \nonumber
\end{eqnarray}
as being the flow density  $\rho$, velocity $\vec v$, pressure $P$ and
current density  $\vec J$. We consider  now a fluid  composed of three
species,  namely electrons  (e), ions  ($i$) and  neutrals  ($n$). The
equations of motion for each specie are
\begin{eqnarray}
\rho_{n} \frac{D\vec{v}_n}{Dt} &=& -\vec{\nabla}P_{n} - \rho_{n}
\vec{\nabla} \Phi_{G} + \vec{F}_{en} + \vec{F}_{in}\\\label{eq:cons_n}
\rho_{i} \frac{D\vec{v}_i}{Dt} &=& -\vec{\nabla}P_{i} - \rho_{i}
\vec{\nabla} \Phi_{G} + \vec{F}_{ei} + \vec{F}_{ni} \notag \\
&&+\ q_i n_i (\vec{E} +\frac{\vec{v}_i}{c} \times \vec{B})\\\label{eq:cons_i}
\rho_{\rm e} \frac{D\vec{v}_{\rm e}}{Dt} &=& -\vec{\nabla}P_{\rm e} - \rho_{\rm e}
\vec{\nabla} \Phi_{G} + \vec{F}_{ie} + \vec{F}_{ne} \notag \\
&&-e\,n_{\rm e} (\vec{E} +\frac{\vec{v}_{\rm e}}{c}\times \vec{B})\label{eq:cons_e}
\end{eqnarray}
where $\Phi_{G}$  is the  gravitational potential and  the collisional
force  of  particles  $\alpha$   on  particles  $\beta$  is  given  by
$\vec{F}_{\alpha \beta} = m_{\alpha\beta} n_{\alpha} \nu_{\alpha\beta}
(\vec{v}_\alpha   -   \vec{v}_\beta)$,  $m_{\alpha\beta}=   m_{\alpha}
m_{\beta}/(m_{\alpha}+m_{\beta})$    being     the    reduced    mass,
$\nu_{\alpha\beta}   =  n_\beta  \mean{\sigma_{\alpha\beta}   v}$  the
collisional frequency and $\mean{\sigma_{\alpha\beta} v}$ the averaged
momentum transfer rate coefficient.

A single  fluid dynamical description  of several species  is relevant
whenever they  are efficiently  collisionally coupled, namely  if they
fulfill $\|\vec{v}_{\alpha=e,n,i}- \vec{v} \| \ll \| \vec{v}\|$. Under
this assumption and  using Newton's principle ($\Sigma_{\alpha, \beta}
\vec  F_{\alpha  \beta}= \vec  0$),  we  get  the usual  MHD  momentum
conservation equation for one fluid
\begin{equation}\label{eq:cons}
\rho  \frac{D\vec{v}}{Dt}   =  -\vec{\nabla}P  -   \rho  \vec{\nabla}
\Phi_{G} + \frac{1}{c}\vec{J} \times \vec{B}
\end{equation}
by adding all  equations for each specie. The  Lorentz force acting on
the mean flow is
\begin{equation}\label{eq:lorentz}
\frac{1}{c}\vec{J} \times \vec{B} = (1 + X) (\vec F_{in} + \vec F_{en}) \ ,
\end{equation}
where $X  = \rho_i/\rho_n$. Even if  the bulk of the  flow is neutral,
collisions with  charged particles give  rise to magnetic  effects. In
turn,  the magnetic  field  is coupled  to  the flow  by the  currents
generated there. This feedback  is provided by the induction equation,
which   requires   the  knowledge   of   the   local  electric   field
$\vec{E}$.  Its expression  is  obtained from  the electrons  momentum
equation
\begin{eqnarray}
\vec{E} + \frac{\vec{v}_{\rm e}}{c} \times \vec{B} & = &
\frac{1}{en_{\rm e}}\big(m_{ie}   n_i\nu_{ie}  \vec{v}_{ie}  +   m_{ne}  n_n
\nu_{ne} \vec{v}_{ne} \big) \nonumber \\
& & - \frac{\vec{\nabla} P_{\rm e}}{e n_{\rm e}}
\end{eqnarray}
where $\vec{v}_{\alpha \beta} \equiv  \vec v_\alpha - \vec v_\beta$ is
the drift  velocity between the  two species. Due to  their negligible
contribution to  the mass  of the bulk  flow, all terms  involving the
electrons  inertia  have  been  neglected (electrons  quite  instantly
adjust themselves to the other forces).

All drift  velocities can be  easily obtained. The  electron-ion drift
velocity is directly provided by  $\vec{v}_{ie} = \vec J/e n_{\rm e}$.
Using  Eq.~(\ref{eq:lorentz}) and noting  that $\rho_{\rm  e} \nu_{en}
\ll \rho_i \nu_{in}$ we get the ion-neutral drift velocity
\begin{equation}
\vec{v}_{in}=\frac{\frac{1}{c}\vec{J}\times\vec{B}}{(1+X)
m_{in} n_i \nu_{in}}  + \frac{m_{en} n_{\rm e} \nu_{en}}{m_{in} n_i \nu_{in}}
\frac{\vec{J}}{ e n_{\rm e}}
\end{equation}
On the same  line of thought, the electrons  velocity is $\vec{v}_{\rm
e} = \vec{v} -(\vec{v}-\vec{v}_{\rm e})$ where
\begin{eqnarray}
\vec{v}   -  \vec{v}_{\rm e}   &=&\frac{\vec{v}_n   -  \vec{v}_{\rm e}}{1+X}  
+ \frac{X}{1+X}(\vec{v}_i -  \vec{v}_{\rm e}) \\
&\simeq& \frac{\vec J}{e n_{\rm e}} - \frac{1}{(1+X)^2} \frac{\frac{1}{c}
\vec{J}\times\vec{B}}{m_{in} n_i\nu_{in}}
\end{eqnarray}
Gathering these  expressions for all  drift velocities, we  obtain the
generalized Ohm's law
\begin{eqnarray}\label{eq:electric}
\vec{E} + \frac{\vec{v} }{c} \times \vec{B}&=& \eta \vec{J}  +
\frac{\frac{1}{c}\vec{J}\times\vec{B}}{e n_{\rm e}} - \frac{\vec{\nabla} P_{\rm e}}{en_{\rm e}} 
\notag \\
&& - \frac{1}{(1+X)^2}
\frac{\frac{1}{c^2}(\vec{J}\times\vec{B})\times\vec{B}}{m_{in} n_i \nu_{in}}
\end{eqnarray}
where $\eta=(m_{ne}n_n \nu_{ne}  + m_{ie} n_i\nu_{ie})/(en_{\rm e})^2$
is the electrical resistivity due to collisions. The corresponding MHD
heating rate writes
\begin{equation}
\Gamma_{\rm MHD}= \vec{J}\cdot\vec{E'} = \vec{J} \cdot \left (\vec{E} +
\frac{\vec{v}}{c} \times \vec{B} \right )
\end{equation}
where $\vec{E'}$ is  the electrical field in the  comoving frame. This
expression leads to equation~(\ref{eq:dissipa}).

The  generalization  of  this  derivation  for a  mixture  of  several
chemical elements  has been done  in a quite straightforward  way. The
bulk flow density becomes $\rho= \overline{\rho_i} + \overline{\rho_n}
+ \rho_{\rm e}$, where the overline stands for a sum over all elements
(ions  and  neutrals),  with  $X=\overline{\rho_i}/\overline{\rho_n}$.
The neutrals and ions velocities  are means over all elements, $\mean{
\vec{v}_{n,i}}          \equiv          \sum_{n,i}          \rho_{n,i}
\vec{v}_{n,i}/\overline{\rho_{n,i}}$.  The  conductivity and collision
terms     are    also     sums    over     all     elements,    namely
$\overline{\eta}=(\overline{m_{ne}n_n\nu_{ne}}    +   \overline{m_{ie}
n_i\nu_{ie}})/(en_{\rm  e})^2$ and  $\overline{m_{in}  n_i \nu_{in}}$,
and are computed using the expressions for the collision frequencies.

\subsection{Momentum transfer rate coefficient}\label{ap:crosssec}

For  ion-electron  collisions  we   use  the  canonical  from
\cite{Schunk75}, summed over all species: 
\begin{eqnarray}
\overline{m_{ie} n_i \nu_{ie}}=m_{\rm e} n_{\rm e} \frac{2\pi e^4}{3 (k_{\text{B}} T_{\rm e})^2}
\sqrt{\frac{8k_{\text{B}} T_{\rm e}}{\pi m_{\rm e}}}\sum_i n_i Z_i^2 \ln{\Lambda_i}
\end{eqnarray}
with       the      Coulomb      factor       $\Lambda_i=(3/2      Z_i
e^3)\sqrt{(k_{\text{B}}T_{\rm e})^3/ \pi n_{\rm e}}$.

For  the  collisions  between   electrons  and  neutrals  we  use  the
expression  of   \cite{Osterbrock61}  for  the   collisional  momentum
transfer rate  coefficient between a  neutral and a  charged particle,
which corrects  the classical one  \citep[eg.,][]{Schunk75} for strong
repulsive forces at close  distances.  Its expression is $\mean{\sigma
v}_{n,i-e}  =  2.41  \pi  e  \sqrt{\alpha_{n}/m_{n,i-e}}$,  where  the
polarizabilities $\alpha_n$  used are  also taken from  Osterbrock. We
thus obtain
\begin{eqnarray}
\overline{m_{en} n_n \nu_{ne}} = 2.41 \pi\, e \,n_{\rm e} \sum_{n={\rm
H,He}} n_n \sqrt{m_{\rm e} \alpha_n}
\end{eqnarray}

Finally, it is mainly the ion-neutral collision momentum transfer rate
coefficient  determines the  ambipolar diffusion  heating.  It  can be
computed  with   the  previous  momentum   transfer  rate  coefficient
expression.   However   as  noted  by   \cite{Draine80}  the  previous
expression  underestimates  $\sigma$  at  high  velocities.  Thus,  as
Draine,  we  take the  ``hard  sphere''  value  for the  cross-section
($\sigma_S  \simeq 10^{-15}$ cm$^2$)  whenever it  is superior  to the
polarizability   one.    For   intermediate   to   hight   ionizations
($f_{\mbox{\tiny  H$^+$}}\gtrsim10^{-4}$)   the  dominant  ion-neutral
collisions are those between  H-H$^+$. Charge exchange effects between
these two species will amplify $\mean{\sigma_{\mbox{\tiny H H$^+$}}v}$
above  the values  expected by  polarizability  alone and  thus it  is
computed separately  (Eqs.~\ref{ap:eq:draine} and \ref{ap:eq:gueiss}).
We thus obtain for ion-neutral collisions
\begin{equation}
\begin{split}
\overline{m_{in} n_i \nu_{in}} &= \frac{1}{2} m_{\rm H}  n_{{\rm H}^+} 
n_{\rm H} \mean{\sigma_{\mbox{\tiny H H$^+$}}v}\\
&+ \sum_{i>H^+} n_i m_{i{\rm H}}
\max{(\sigma_S\,\tilde{v}; \mean{\sigma  v}_{\text{H,i}})}\\
&+ \sum_{i} n_i m_{i{\rm He}} \max{(\sigma_S\,\tilde{v};
\mean{\sigma v}_{\text{He},i})}
\end{split}
\end{equation}
where  $\tilde{v}=\sqrt{8k_{\text{B}}T/\pi  m_{in}  + v_{in}^2}$.  For
$\tilde{v}<1000$ km s$^{-1}$.

The value of $\mean{\sigma_{\mbox{ \tiny H H$^+$}}v}$ which we used is
given by \cite{Draine80},
\begin{eqnarray}\label{ap:eq:draine}
\frac{\mean{\sigma_{ \mbox{{\tiny H\,H$^+$}}}v}}{1 \,\text{cm}^3\text{s}^{-1}} \simeq
\begin{cases}
3.26\times10^{-9}& \tilde{v}<2\:\mbox{km  s}^{-1}\\
2.0\times10^{-9}\,\tilde{v}^{0.73}&\tilde{v}\geq 2 \:\mbox{km  s}^{-1}
\end{cases}
\end{eqnarray}
\cite{Safier93a}  used  the  expression $\tilde{v}=v_{in}$  which,  as
discussed in section~4.8, results  in a smaller momentum transfer rate
coefficient.  \cite{GeissBuergi86} computed  another expression of the
H-H$^+$ momentum transfer rate coefficient, which provides
\begin{equation}\label{ap:eq:gueiss}
\begin{split}
\mean{\sigma_{   \mbox{{\tiny    H\,H$^+$}}}v}   &=   1.12\times10^{-8}
\:T_4^{\frac{1}{2}} (1-0.12\log{T_4})^2\\
+\ \big(2.4&-0.34 (1+2\log{T_4})^2\big)\times10^{-9}\:\: \text{cm}^3\text{s}^{-1}
\end{split}
\end{equation}
In  Figure~\ref{fig:cross}  we compared  both  momentum transfer  rate
coefficients, they typically  differ in 40\%, which can  be used as an
estimate of their accuracy. It  is thus the uncertainty in the H-H$^+$
momentum transfer rate coefficient  that dominates the final intrinsic
uncertainty of our calculations.

\section{Dust implementation}\label{ap:dust}

\begin{figure} \label{ap:fig:dust}%-----------------------------------
  \begin{center}
    \resizebox{6cm}{!}{\includegraphics{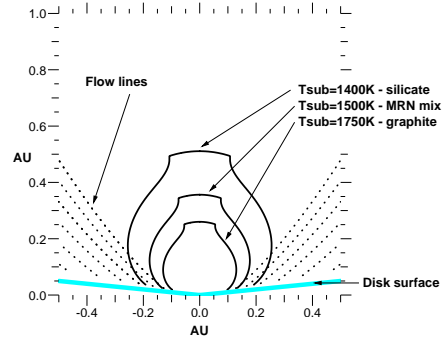}}
  \end{center}
  \caption{ Dust sublimation surfaces geometry for the adopted radiation
    field.} 
\end{figure} %--------------------------------------------------------

As shown by  \citeauthor{Safier93a} if there is dust  in the disk, the
wind is  powerful enough to drag  it along. Thus disk  winds are dusty
winds. Dust is  important for the wind thermal  structure mainly as an
opacity  source  affecting the  photoionisation  heating  at the  wind
base. To  compute the dust opacity  we need a description  of its size
distribution,  its wavelength  dependent absorption  cross-section and
the inner  dust sublimation surface. In  the inner flow  zones and for
high accretion rates  the strong stellar and boundary  layer flux will
sublimate  the   dust,  creating  a   dust  free  inner   cavity  (see
figure~\ref{ap:fig:dust}).   Results  on  the  evolution  of  dust  in
accretion disks by \cite{Schmitt97} show  that at the disk surface the
initial  dust  distribution isn't  much  affected  by coagulation  and
sedimentation  effects.   Thus  we  assume  a  MRN  dust  distribution
\citep{Mathis77, DraineLee84}:
\begin{equation}
dn_i=n_{\rm H}\,A_i\,a^{-3.5}\,da
\end{equation}
where $dn_i$  is the number  of particles of type  $i$ (``astronomical
silicate''  -- Sil or  graphite --  C) with  sizes in  $[a,a+da]$, and
$0.005    \mbox{$\mu$m}    \leq    a   \leq    0.25    \mbox{$\mu$m}$,
$A_{Sil}=10^{-25.11}$  cm$^{2.5}$   H$^{-1}$  and  $A_{C}=10^{-25.16}$
cm$^{2.5}$ H$^{-1}$.  We then  proceed by averaging all relevant grain
quantities function of size and species ($F_i(a)$) by the size/species
distribution,
\begin{equation}\label{ap:eq:dustd}
\mean{F_i(a)}_a   =  \int_{a_{\rm  min}}^{a_{\rm   max}}  \sum_{i={\rm
Sil,C}} F_i(a) \, \frac{dn_i}{N_T}
\end{equation}

In order  to compute  the sublimation radius  some description  of the
dust temperature must  be made. For simplicity, we  assume the dust to
be in thermodynamic equilibrium with the radiation field, the dominant
dust  heating mechanism.  In our  case, the  central  source radiation
field  will  dominate  throughout  the  jet, except  probably  in  the
recollimation  zone,  where  the  strong gas  emission  overcomes  the
central  diluted field.  However  in  this region  dust  is no  longer
relevant  for  the  gas  thermodynamics  and we  will  therefore  only
consider  dust heating by  the central  source.  The  dust temperature
$T_{\rm  gr}$ for  a grain  of size  $a$ is  obtained by  equating the
absorbed to the emitted radiation \cite[eg.,][]{TielensHollenbach85},
\begin{eqnarray} \label{eq:dust}
4 \pi a^2 \mean{Q_a}(T_{\rm gr}) \sigma T_{\rm gr}^4 = \pi a^2 \,\int_0^\infty
Q_a^{\rm abs}(\nu) 4\pi J_\nu d\nu
\end{eqnarray}
where  $a$   is  the  grain  size  $\mean{Q_a}(T_{\rm   gr})$  is  the
Planck-averaged     emissivity    \citep{DraineLee84,    LaorDraine93,
DraineMal93},  $\sigma$ is  the  Stefan-Boltzmann constant,  $Q_a^{\rm
abs}(\nu)$  is   the  dust  absorption   efficiency\footnote{The  dust
absorption efficiency is related  to the dust absorption cross-section
by $\sigma_{a,\nu}^{\rm abs}=\pi a ^2 Q_a^{\rm abs}(\nu)$.}  and $4\pi
J_\nu$  is the  central source  radiation flux  at the  grain position
given by equation \ref{eq:rad}. Averaging out the previous equation by
the size/species distribution (eq.~\ref{ap:eq:dustd}) we obtain,
\begin{equation}
4    \mean{Q_a^{\rm em}(T_{\rm gr})}   \sigma    T_{\rm gr}^4    =   \int_0^\infty
\mean{Q_a^{\rm abs}}(\nu) F(\nu) e^{-\tau_\nu(r,\theta)} d\nu
\end{equation}
where  we  describe the  central  source  flux  by $F(\nu)$  which  is
attenuated only by the dust opacity $\tau$. For simplicity $F(\nu)$ is
taken  as  exactly  the  same  as  in  \citeauthor{Safier93a},  ie.  a
classical boundary layer \citep{Bertout88}.  The sublimation radius is
obtained from the  previous expression by noting that  at its position
$\tau_\nu=0$,
\begin{equation}
\frac{\mean{r_{\rm sub}(\theta)}}{R_\ast}=
\sqrt{\frac{g_\ast(\theta)\,\mean{Q_a^{\rm abs}(T_\ast)}T_\ast^4+
g_{\rm bl}(\theta)\,
\mean{Q_a^{\rm abs}(T_{\rm bl})} T_{\rm bl}^4 }{4 \mean{Q_a^{\rm em}(T_{\rm sub})} T_{\rm sub}^4 }}
\end{equation}
where  $T_{\rm bl}$  and $T_{\ast}$  are the  boundary layer  and star
temperatures,    $R_\ast$    the    stellar   radius    and    $g_{\rm
bl}(\theta)$/$g_\ast(\theta)$ are the  $\theta$ dependent terms of the
radiation field (given in  Bertout et al. and \citeauthor{Safier93a}).
We assume a dust sublimation temperature $T_{\rm sub}$ of 1500~K.

With the dust sublimation radius in hand we can now proceed to compute
the dust optical depth defined as,
\begin{equation} \label{eq:dust_tau}
\tau_\nu(r,\theta)  =  \int_{r_{\rm  in}(\theta)}^{r}  \kappa_\nu^{\rm
abs}(r')dr'  =  \int_{r_{\rm  in}(\theta)}^{r} n_{\rm  H}  (r',\theta)
\overline{\sigma(\nu)_a} dr'
\end{equation}
where  $r_{\rm in}(\theta)$  is the  radius inside  which there  is no
dust.  This  radius is  given by the  inner flow  line $r_{\varpi_{\rm
i}}(\theta)$   and    by   the   sublimation    radius   $\mean{r_{\rm
sub}(\theta)}$   (see  figure~\ref{ap:fig:dust})  such   that  $r_{in}
(\theta) =  \text{max} (\mean{r_{\rm sub}};r_{\varpi_{\rm  i}})$.  The
dust absorption cross-section ($\overline{\sigma(\nu)_a}$) is,
\begin{eqnarray}
\overline{\sigma(\nu)_a}=\int_{a_{\rm min}}^{a_{\rm max}}  & & \pi a^2
\big[ Q^{\rm abs}_{\rm Sil}(a,\nu) A_{\rm Sil} \nonumber \\ &+ & Q^{\rm
abs}_{\rm C}(a,\nu) A_{\rm C} \big] a^{-3.5} da
\end{eqnarray}
Using the  self-similarity of $n_{\rm H} (r,\theta)$  we can integrate
equation~\ref{eq:dust_tau} to obtain,
\begin{equation}
\tau_\nu(r,\theta)  =n_{\rm H}(r,\theta)\,\overline{\sigma(\nu)} \,
2r\Big(\sqrt{\frac{r}{\mean{r_{\rm in}(\theta)}}}-1\Big)
\end{equation}
which  was  used  in  equation~\ref{eq:rad}.  We note  that  at  large
distances from  the source,  the optical depth  converges to  a finite
value,  proportional  to  $\dot{M}_{\text{acc}}$  and  whose  $\theta$
variation is  function of the  self similar wind solution  and central
source radiation  field.  Thus for high accretion  rates, although the
central source  radiation hardens, the  {\em outer} zones of  the wind
base are less photoionized than for smaller accretion rates.

\section{Consistency checks}\label{ap:cons}

\subsection{Dynamical assumptions}\label{ap:cons:dyn}

%______________________________________________________________________ 
\begin{figure}
\begin{center}
   \resizebox{8cm}{!}{\includegraphics{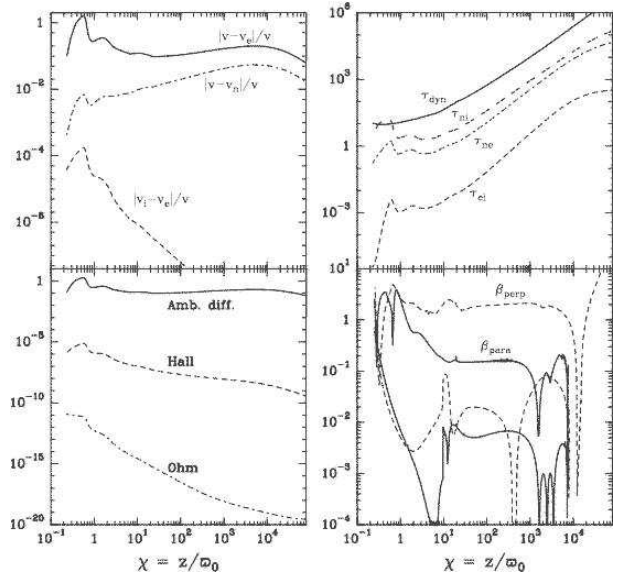}}
\end{center}
\caption{ {\bf  Top  left:} We plot the relevant drift speeds normalized to 
the fluid poloidal velocity. The worst case of one fluid approximation
violation  is obtained  for model  C, $\dot{M}_{\rm  acc}=10^{-8} {\rm
M}_\odot  \text{yr}^{-1}$ and  $\varpi_0=1$ AU.   {\bf Top  right:} We
plot  the  $\tau_{\alpha  \beta}$  versus dynamical  $\tau_{\rm  dyn}$
time-scales (s)  versus $\chi$, normalized to the  Keplerian period at
the line footpoint (for a 1  M$_\odot$ star).  We plot only the longer
time-scales ($\tau_{\rm in} =  \tau_{\rm ni}/f_i$ and $\tau_{\rm en} =
\tau_{\rm ne}/f_{\rm e}$).  The {\em worst} case is obtained again for
model C, $\dot{M}_{\rm  acc}=10^{-8} {\rm M}_\odot \text{yr}^{-1}$ and
$\varpi_0=1$  AU.  {\bf  Bottom left:}  Ideal MHD  tests for  the {\em
worst}  situation (model C,  $\dot{M}_{\rm acc}=10^{-8}  {\rm M}_\odot
\text{yr}^{-1}$  and $\varpi_0=1$  AU).  As  expected from  our heated
winds, the ambipolar diffusion term  is the dominant one.  {\bf Bottom
right:} ratios of  the thermal pressure gradient to  the Lorentz force
versus   $\chi$,   along   ($\beta_\parallel$,   solid)   and   across
($\beta_\perp$, dash)  a magnetic surface anchored  at $\varpi_0$. The
{\em  worst}  case for  $\beta_\parallel$  is  obtained  for model  A,
$\varpi_0=   1$   AU,   $\dot{M}_{\rm   acc}=10^{-8}   {\rm   M}_\odot
\text{yr}^{-1}$, the  {\em best} for  model A, $\varpi_0= 0.1$  AU and
$\dot{M}_{\rm acc}=10^{-5} {\rm M}_\odot \text{yr}^{-1}$. With respect
to $\beta_\perp$, the  {\em worst} case is for  model C, $\varpi_0= 1$
AU and $\dot{M}_{\rm acc}=10^{-8} {\rm M}_\odot \text{yr}^{-1}$, while
the {\em  best} is for model  A, $\varpi_0= 0.1$  AU and $\dot{M}_{\rm
acc}=10^{-5}  {\rm M}_\odot  \text{yr}^{-1}$. Although  definitely not
negligible in  some models, those compatible with  observations do not
show important deviations from the ``cold'' jet approximation.}
\label{fig:cons_dyn}%
\end{figure}
%
%______________________________________________________________________ 

First,  local charge neutrality  is always  achieved. For  example, we
achieve a  maximum Debye length of $r_{\text{D}}\sim10^5  {\rm cm}$ at
the  outer   radius  of  the  recollimation  zone   (model  C,  lowest
$\dot{M}_{\rm acc}$).

Second,  single fluid  approximation requires  that  relative velocity
drifts of all species ($\alpha=$ ions, electrons, neutrals) $\|\vec{v}
- \vec{v}_{\alpha}\|  /  \|\vec{v}\|$ are  smaller  than unity.  These
drifts are higher for lower accretion rates and at the outer wind base
(due    to   the    decrease    in   density    and   velocity,    see
Eq.~\ref{eq:scalings}).  In figure~\ref{fig:cons_dyn}  we  present the
{\em worst} case  for the drift velocities, showing  that our jets can
be indeed approximated by single fluid calculations.

We assumed  gas thermalization, which is achieved  only if collisional
time-scales   between  species  $\tau_{\alpha\beta}   =  1/\nu_{\alpha
\beta}$   are    much   smaller   than    the   dynamical   time-scale
$\tau_{\text{dyn}} =  \varpi_0 (d v_z/d \chi)^{-1}$.  In the collision
network  considered here,  the longer  time-scales  involve collisions
with   neutrals.   However,  even   in   the   worst  situation   (see
figure~\ref{fig:cons_dyn}),   after   the   wind  base   they   remain
comfortably below the above dynamical time-scale.
 
Our  dynamical  jet  solutions  were  derived  within  the  ideal  MHD
framework. This assumption  requires that all terms in  the right hand
side  of the  generalized Ohm's  law  (equation~\ref{eq:electric}) are
negligible when  compared to  the electromotive field  $\vec{v} \times
\vec{B}/c$. We consider Ohm's  term $\|\eta \vec{J} \|$, Hall's effect
$\|\vec{J}\times\vec{B}\|/c\,e n_{\rm e}$  and the ambipolar diffusion
term   $(\frac{\overline{\rho_n}}{\rho})^2  \|(\vec{J}  \times\vec{B})
\times\vec{B}\|/c^2  \overline{m_{in} n_i  \nu_{in}}$ (effects  due to
the  electronic pressure gradient  are small  compared to  the Lorentz
force ---  Hall's term ---).  In  figure~\ref{fig:cons_dyn} we present
the {\em worst} case for our ideal MHD checks. We find that deviations
from ideal  MHD remain negligible,  despite the presence  of ambipolar
diffusion.  As expected, this is the dominant diffusion process in our
(non  turbulent)  MHD jets.  Ambipolar  diffusion  is  larger for  low
accretion rates and at the outer wind {\em base} (because the ratio of
the    ambipolar    to    the    electromotive    term    scales    as
$(\dot{M}_{\text{acc}}\,f_i)^{-1}$.

The worst case  for the previous three tests is,  as expected, for the
model  that attains  the  lowest  density: Model  C,  with the  lowest
accretion   rate  (\mbox{$\dot{M}_{\rm   acc}=10^{-8}\,   {\rm  M}_\odot
\text{yr}^{-1}$}) and  at the outer  edge footpoint (\mbox{$\varpi_0=1\,
{\rm AU}$}).

The  dynamical  jet  evolution  was calculated  under  the  additional
assumption of negligible thermal  pressure gradient (cold jets). Since
it is the gradient that provides  a force, one should not just measure
(along one field line) the  relative importance of the gas pressure to
the     magnetic     pressure     (usual     $\beta=     P/(B^2/8\pi)$
parameter). Instead,  we compare the thermal pressure  gradient to the
Lorentz    force,   along   ($\beta_\parallel$)    and   perpendicular
($\beta_\perp$) to the flow, namely
\begin{eqnarray}
\beta_\parallel &=& \frac{\nabla_\parallel P}{F_\parallel}= c
\frac{\vec{v}_{\rm p} \cdot \nabla P}{ \vec{v}_{\rm p} \cdot (\vec J \times \vec B)} \\
\beta_\perp &=& \frac{\nabla_\perp P}{F_\perp} = c \frac{\nabla a \cdot
  \nabla P}{ \nabla a \cdot (\vec J \times \vec B)} 
\end{eqnarray}
Here  $a(\varpi, z)$  is the  poloidal magnetic  flux  function, hence
$\nabla  a$ is  perpendicular to  a magnetic  surface. High  values of
$\beta_\parallel$  imply that  the thermal  pressure gradient  plays a
role in  gas {\em acceleration}, whereas high  values of $\beta_\perp$
show that it affects the gas {\em collimation}.

In  figure~\ref{fig:cons_dyn} we  plot the  worst case  of  cold fluid
violation and best case of  cold fluid validity.  Again the worst case
appears at lower  accretion rates and in outer wind  zones.  It can be
seen that  high values of  $\beta_\perp$ and $\beta_\parallel$  can be
attained, hinting  at the  importance of gas  heating on  jet dynamics
(providing  both enthalpy  at  the  base of  the  jet and/or  pressure
support against recollimation further  out).  We underline that models
inconsistent with the cold fluid approximation are those found to have
the    largest     difficulty    in    meeting     the    observations
\citepalias{Garcia2001b}.   Conversely, models  that  better reproduce
observations  also fulfill  the  cold fluid  approximation. For  those
models, the thermal pressure  gradient appears to be fairly negligible
with respect to the Lorentz force.

\subsection{Thermal assumptions} \label{ss:ign_h}
%______________________________________________________________________ 
\begin{figure}
\begin{center}
   \resizebox{9cm}{!}{\rotatebox{-90}{\includegraphics{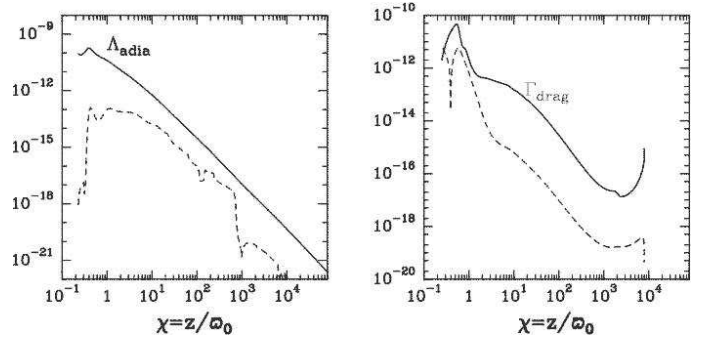}}}
\end{center}
\caption{Ignored heating/cooling terms (in erg s$^{-1}$ cm$^{-3}$). 
{\bf Left:}  We compare the  cooling term $- \frac{3}{2}k  \tilde{n} T
Df_{\rm  e}/Dt$ (dashed) with  adiabatic cooling  $\Lambda_{\rm adia}$
(solid)  for  the  {\em  worst}  case  (model  C,  \mbox{$\dot{M}_{\rm
acc}=10^{-8}  {\rm M}_\odot  \text{yr}^{-1}$} and  $\varpi_0=0.1$ AU).
{\bf Right:}  Grain heating/cooling rate  $|\Gamma_{\rm gr}|$ (dashed)
compared with ambipolar  diffusion heating $\Gamma_{\rm drag}$ (solid)
for  the {\em worst}  case (model  A, $\dot{M}_{\rm  acc}=10^{-5} {\rm
M}_\odot \text{yr}^{-1}$  and $\varpi_0=1$~AU).  $\Gamma_{\rm  gr}$ is
only significant at the very base of the wind, and will not affect the
thermal state further out in the jet.}
\label{fig:cons_the}%
\end{figure}
%
%______________________________________________________________________ 
 
Finally, we check that all ignored heating/cooling  processes are not
relevant when compared to adiabatic cooling and ambipolar diffusion
heating.

The  first ignored  process is  the term  $- \frac{3}{2}k  \tilde{n} T
Df_{\rm e}/Dt$.  This term decreases for increasing accretion rate and
$\varpi_0$ due to the lower  ionizations found in these regions. It is
plotted  in  figure~\ref{fig:cons_the} for  the  worst case  (model~C,
\mbox{$\dot{M}_{\rm  acc}=10^{-8} {\rm  M}_\odot  \text{yr}^{-1}$} and
$\varpi_0=0.1$  AU).    There,  it  reaches  at   most  $\sim$13\%  of
$\Lambda_{\rm adia}$.   Typical values for higher  accretion rates are
only $\simeq$ 0.1\% of $\Lambda_{\rm adia}$.

Next we  consider heating/cooling  of the gas  by collision  with dust
grains, given by \cite{HollenbachMcKee79}:
\begin{equation}
  \Gamma_{\text{gr}} = n \mean{n_{\text{gr}} \sigma_{\text{gr}} } 
  \sqrt{ \frac{ 8 k_{\text{B}} T }{\pi m_{\text{H}} } } f(2k_{\text{B}} T_{\rm gr}
  - 2k_{\text{B}} T) 
\end{equation}
where $\mean{n_{\text{gr}}  \sigma_{\text{gr}}}$ is computed  from the
adopted MRN distribution, and  $f=0.16$ is the sticking parameter that
takes into  account charge  and accommodation effects  for a  warm gas
\citep{HollenbachMcKee79}.     With    these    values    the    grain
heating/cooling becomes,
\begin{equation}
  \Gamma_{\text{gr}} = 4.78 \times 10^{-34} n^2 (T_{\text{gr}}-T) \hspace{.5cm}
  \mbox{erg s$^{-1}$ cm$^{-3}$}
\end{equation}
This term increases with  increasing $\varpi_0$ and accretion rate. In
figure~\ref{fig:cons_the}  we  compare  it  (in absolute  value)  with
$\Gamma_{\rm drag}$  in the  case where its  contribution is  the most
important   (model  A,   $\dot{M}_{\rm   acc}=10^{-5}  {\rm   M}_\odot
\mbox{yr}^{-1}$ and  $\varpi_0=1$~AU). $\Gamma_{\rm gr}$  is initially
positive (dust hotter than the  gas), but changes sign at $\chi \simeq
0.4$,  where  the  gas  becomes  hotter than  the  dust,  becoming  an
effective cooling term. It is only significant at the very base of the
wind,  where it  exceeds the  ambipolar  diffusion heating  term by  a
factor  $\simeq$ 3.5.  However,  this effect  will not  have important
consequences in  terms of observational  predictions: We will  show in
next  section  that  the  thermal  state in  the  hot  plateau  (where
forbidden line  emission is excited)  is not sensitive to  the initial
temperature.  Furthermore,  the outer streamlines  at $\varpi_0 \simeq
1$~AU contribute much  less to the line emission  than the inner ones.
At lower  accretion rates $\le 10^{-6}  {\rm M}_\odot \mbox{yr}^{-1}$,
$\Gamma_{\rm gr}$ is always $\le$10 \% of $\Gamma_{\rm drag}$.

Heating  due to cosmic  rays, which  could be  important in  the outer
tenuous zones of the wind is \citep{SpitzerTomasko68},
\begin{eqnarray}
\Gamma_{cr} = n_{\rm H}  \zeta \Delta E = 1.9\times 10^{-28}  \, n_{H}
\mbox{ erg s$^{-1}$ cm$^{-3}$}
\end{eqnarray}
where $\zeta$ is the ionization rate which we took as $\zeta=3.5\times
10^{-17}$ s$^{-1}$  \citep{Webber98} and  $\Delta E =  3.4$ eV  is the
average  thermal energy  transmitted  to the  gas  by each  ionization
\citep{SpitzerTomasko68}.  This  effect  is  at most  $\sim  3.6\times
10^{-6}$  times  $\Gamma_{\text{drag}}$  for  model  A,  $\dot{M}_{\rm
acc}=10^{-5} {\rm M}_\odot \text{yr}^{-1}$ and $\varpi_0=0.1$ AU.
 
Finally  the  thermal  conductivity  along magnetic  field  lines  was
computed  with  the  Spitzer  conductivity  for a  fully  ionized  gas
\citep{Lang99}  and  is  irrelevant  (at  most  $\sim10^{-6}$  of  the
adiabatic   cooling  term)   the   maximum  being   achieved  at   the
recollimation zone  where the physical  validity of our  MHD solutions
ends.   It should be  pointed out  that \cite{NowakUlm77}  compute the
thermal  conductivity for  a partially  ionized mixture  in ionization
equilibrium and  found that for low  temperatures $T\sim10^{3-4}$K the
Spitzer  expression  underestimates   the  conductivity  by  a  factor
$10^2$. However this is still too small to be important.

\subsection{Dependence on initial conditions} \label{sss:t_init}
%______________________________________________________________________ 

\begin{figure}
   \resizebox{8cm}{!}{\includegraphics{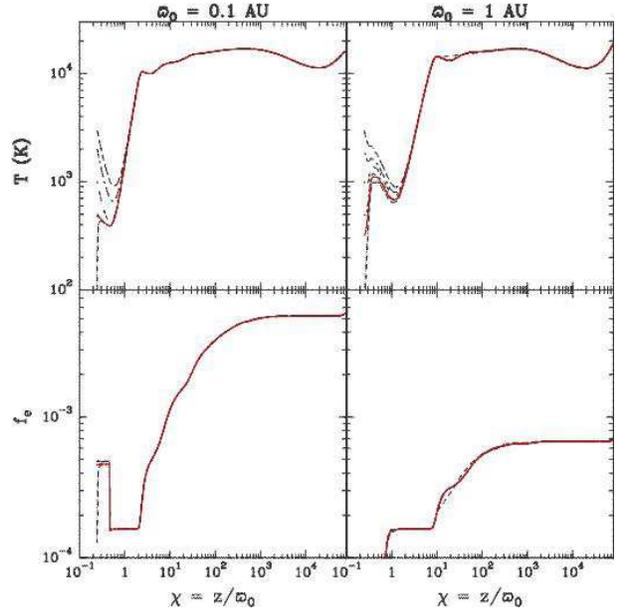}}
\caption{Effect of initial temperature on the thermal evolution
for model B,  $\dot{M}_{\rm acc}=10^{-6}{\rm M}_\odot \text{yr}^{-1}$.
The  several initial temperatures  are in  dashed 50~K,  100~K, 500~K,
1000~K, 2000~K and  3000~K. In solid we plot  the solution obtained by
our  standard  initial  conditions.   Note that  the  almost  vertical
evolution of the temperature for  very low initial temperatures is not
an artifact. The field line  anchored at $\varpi_0=0.1$ AU crosses the
sublimation surface at $\chi\simeq 0.5$.}
\label{fig:t_init}%
\end{figure}
%
%______________________________________________________________________ 

Formally, our temperature integration  is an initial value problem. In
the   absence   of  a   self-consistent   description   of  the   disc
thermodynamics,  there  is some  freedom  in  the initial  temperature
determination. It  is therefore crucial  to check that  the subsequent
thermal  evolution of  the  wind  does not  depend  critically on  the
adopted initial value.

\citeauthor{Safier93a}  obtained the  initial temperature  by assuming
the poloidal  velocity at the slow magnetosonic  point ($v_{\rm p,s}$)
to be  the sound speed for  adiabatic perturbations $ T_{\rm  s} = \mu
m_{\rm H}  v_{\rm p,s}^2/ \gamma k_{\text{B}}$.  Here,  we have chosen
to compute the initial  temperature assuming local thermal equilibrium
\mbox{$DT/Dt=0$}. Our method  produces lower initial temperatures than
\citeauthor{Safier93a} due to adiabatic cooling.

For high accretion rates $\dot{M}_{\rm acc}\,\geq 10^{-6}{\rm M}_\odot
\text{yr}^{-1}$  our  initial  temperature  versus  $\varpi_0$  has  a
minimum at  the beginning  of the dusty  zone: Inside  the sublimation
cavity, the thermal equilibrium is between photoionization heating and
adiabatic   cooling.   Just  beyond   the  dust   sublimation  radius,
photoionization  heating  is  strongly  reduced,  but  the  ionization
fraction is still too high  for efficient drag heating, resulting in a
low initial equilibrium temperature.

The initial  ionization fraction  is similarly determined  by assuming
local  ionization equilibrium  $Df_A^i/Dt =  0$ for  all  elements. It
decreases with $\varpi_0$. For $\dot{M}_{\rm acc}=10^{-5}{\rm M}_\odot
\text{yr}^{-1}$  and $\varpi_0  \geq 0.8$  AU, the  initial ionization
fraction  is set  to a  minimum  value by  assuming that  Na is  fully
ionized, which  is somewhat arbitrary.   However, as gas is  lifted up
above the disk plane, the dust opacity decreases and the gas heats up,
so  that  ionization becomes  dominated  by  other photoionized  heavy
species and by protons, all computed self consistently.

In  order to  check that  our  results do  not depend  on the  initial
temperature,  we  have  run model  B  for  a  broad range  in  initial
temperatures.  As shown in  Figure~\ref{fig:t_init}), we find that the
thermal and  ionization evolution  quickly becomes insensitive  to the
initial temperature.   If we start  with a temperature lower  than the
local isothermal condition, the dominant adiabatic cooling is strongly
reduced,  and the  gas strongly  heats up,  quickly converging  to our
nominal  curve.   If  we  start  with a  higher  initial  temperature,
adiabatic cooling is  stronger, and we have the  characteristic dip in
the temperature found by \citeauthor{Safier93a}. Our choice of initial
temperature has the advantage of  reducing this dip, which is somewhat
artificial (see figure~\ref{fig:t_init}).  In either case, we conclude
that  our results are  robust with  respect to  the choice  of initial
temperature. In particular,  the distance at which the  hot plateau is
reached, which  has a crucial  effect on line profile  predictions, is
unaffected.

\bibliographystyle{apj}
\bibliography{references}

\end{document}